\documentclass{IET}

\usepackage{amssymb}
\usepackage{amsthm}
\usepackage[scaled=0.92]{helvet}
\usepackage{courier}
\usepackage{graphicx} 
\usepackage{amsmath}
\usepackage{bm}
\usepackage{amsfonts}
\usepackage{mathrsfs}
\usepackage{enumerate}
\usepackage{subfigure}
\usepackage{hyphenat} 

\DeclareMathOperator*{\argmin}{arg\,min}

\begin{document}

\title{Filter-based regularisation for impulse response modelling
\footnote{This paper is a postprint of a paper submitted to and accepted for publication in IET Control Theory and Applications and is subject to Institution of Engineering and Technology Copyright. The copy of record is available at IET Digital Library, under the DOI: 10.1049/iet-cta.2016.0908.}}

\author[1,**]{Anna Marconato}

\author{Maarten Schoukens}

\author{Johan Schoukens}
\affil{Dept. ELEC, Vrije Universiteit Brussel, Pleinlaan 2, 1050 Brussels, Belgium}
\affil[**]{anna.marconato@vub.ac.be}

\abstract{In the last years, the success of kernel-based regularisation techniques in solving impulse response modelling tasks has revived the interest on linear system identification. In this work, an alternative perspective on the same problem is introduced. Instead of relying on a Bayesian framework to include assumptions about the system in the definition of the covariance matrix of the parameters, here the prior knowledge is injected at the cost function level. The key idea is to define the regularisation matrix as a filtering operation on the parameters, which allows for a more intuitive formulation of the problem from an engineering point of view. Moreover, this results in a unified framework to model low-pass, band-pass and high-pass systems, and systems with one or more resonances. The proposed filter-based approach outperforms the existing regularisation method based on the TC and DC kernels, as illustrated by means of Monte Carlo simulations on several linear modelling examples.}

\maketitle

\section{Introduction}
\label{sec1}

The identification of linear time-invariant systems has been extensively studied in the last decades, and has been considered a `solved' problem for several years \cite{Lju99}, \cite{PinSch12}. Nevertheless, recent developments on the use of regularisation techniques for impulse response modelling have shed new light on this old problem. In particular, it has been shown that the standard prediction error method/maximum likelihood (PEM/ML) approaches can be outperformed by introducing a clever way to add prior information in the estimation problem \cite{Pil10}, \cite{Chen12}. 

More in details, robust kernel-based regularisation methods for impulse response estimation have been recently designed relying on the theory of Gaussian processes \cite{Ras05}. Exploiting the Bayesian framework, information about the system properties is included in the definition of the covariance matrix of the parameters, see \cite{Pil14} for a recent survey. Typical examples of such properties are the smoothness and exponential decay of the impulse response, which result in the definition of the widely-used DC and TC kernels (the latter are also known as first-order stable spline kernels) \cite{Chen12}.

The objective of the present work is to study regularised impulse response modelling from a different perspective. Instead of designing the regularisation term starting from the covariance matrix of the parameters, the estimation problem is analysed directly at the cost function level, allowing for an intuitive interpretation of regularisation problems from an engineering point of view. The core idea is to regard the regularisation term in the cost function as a filtering operation on the parameters to be estimated. In practice, this results in a definition of the regularisation matrix that includes the properties that one wants to penalise, i.e.\ the inverse of the assumed system properties.

In this way, by exploiting this filter-based approach it is possible to design a flexible algorithm to model, in a unified framework, low-pass, band-pass and high-pass systems, and systems with one or multiple resonances.

Recently, examples of more general kernel structures have been introduced in kernel-based identification to model systems characterised by complicated dynamics, e.g.\ based on stochastic state space models \cite{ChenIFAC14} or orthonormal basis functions \cite{ChenECC15}. A first step in the direction of extending basic kernel structures was already taken in \cite{Pil11}, where a finite-dimensional parametric component was added to the stable spline kernel to include high-frequency poles in the impulse response representation. In this paper, starting from an alternative interpretation of the problem, we further increase the flexibility of the regularisation approach, by defining a general framework in which the user can incorporate different prior assumptions in a natural way. Furthermore, thanks to this dual framework, one can hopefully achieve a better understanding of regularisation methods in system identification, and explain the reasons for their success.


This paper extends the results presented in \cite{AnnaECC16}. In the present work, the filter interpretation ideas are further developed and several different kernels based on the smoothness and exponential decay properties are analysed in depth (Section~\ref{sec4}), while in \cite{AnnaECC16} only the simple TC kernel case was discussed. Moreover, the effectiveness of the proposed method is tested on a wider range of systems since three, more challenging, examples of systems with one or two resonances are also considered now (Section~\ref{sec6}).

The rest of the paper is organised as follows. The classic and the regularised approach to impulse response modelling are presented in Section~\ref{sec2}. Section~\ref{sec3} deals with the different ways of including information about the system in the regularisation problem. The filter interpretation idea is further investigated in Section~\ref{sec4}. The details of the filter-based approach for regularised impulse response modelling are explained in Section~\ref{sec5}, and the results obtained on different Monte Carlo simulation examples are discussed in Section~\ref{sec6}. Concluding remarks end the paper in Section~\ref{concl}.


\section{Problem formulation}
\label{sec2}

In this work, we consider the estimation of finite impulse response (FIR) models for linear time-invariant systems in a discrete-time setting, based on a set of input--output data $\left\{(u(t),y(t))\right\}_{t=1}^N$. The output data are assumed to be corrupted by additive white Gaussian noise, characterised by zero mean and variance $\sigma^2$, and independent from the input signal. 

The order $n$ of the FIR model is considered to be fixed, and the modelled output can then be written as:
\begin{equation}
\label{FIReq}
\hat{y}(t)=\sum_{k=0}^{n-1} g_k u(t-k),
\end{equation}
where $g_k$ are the $n$ impulse response coefficients to be estimated.

\subsection{Classic approach}

The standard PEM/ML problem formulation is given by:
\begin{equation}
\label{PEM}
\hat{\theta}=\argmin_\theta \sum_{t=1}^{N} (y(t)-\hat{y}(t,\theta))^2,
\end{equation}
where $\theta \in \mathbb{R}^n$ contains all impulse response coefficients $g_k$, $k=0,\ldots,n-1$.

Solving (\ref{PEM}) results in the least squares estimate:
\begin{equation}
\label{LSsol}
\hat{\theta}_{\text{ls}}=\argmin_\theta \|Y-\Phi\theta\|^2 = (\Phi^T\Phi)^{-1}\Phi^T Y.
\end{equation}

Note that a compact notation is preferred here, such that all output data $y(t)$, $t=1,\ldots,N$ are collected in the column vector $Y$, and the regressor matrix $\Phi$ contains shifted instances of the input data $u(t)$:
\begin{equation}
\label{Phieq}
\Phi =
  \begin{pmatrix}
   u(1) & u(0) & \cdots & u(-n+2) \\
   u(2) & u(1) & \cdots & u(-n+3) \\
   \vdots  & \vdots  & \ddots & \vdots  \\
   u(N) & u(N-1) & \cdots & u(N-n+1)
  \end{pmatrix}.
\end{equation}
Here, the initial conditions (input samples for $t=-n+2,\ldots,0$) are assumed to be equal to $0$ for simplicity.

\subsection{Regularised approach}

In regularisation methods, a penalty term on the model complexity is included in the cost function, to decrease the variance on the estimated parameters (this comes at the expense of introducing in the model error a bias component, which is typically quite small). This results in the regularised estimate:
\begin{equation}
\label{regsolution}
\hat\theta_{\text{reg}}=\argmin_\theta \|Y-\Phi\theta\|^2+\theta^TR\theta=(\Phi^T\Phi+R)^{-1}\Phi^TY.
\end{equation}
The regularisation matrix $R$ is a symmetric positive-semidefinite matrix, which is introduced to impose a different complexity penalty for the parameters in $\theta$. A way of doing this, as will be explained in the next section, is to incorporate in $R$ prior knowledge about the underlying system.

\section{Two alternative ways of incorporating prior information in the problem}
\label{sec3}

%
%

\subsection{Kernel-based problem formulation}
\label{secker}

In order to include prior information in the regularisation problem, one can introduce the covariance matrix $P$ of the parameters $\theta$, also known as the \textit{kernel} matrix, and define \cite{Chen12}:
\begin{equation}
\label{Rinv}
R=\sigma^2P^{-1}.
\end{equation}
This is equivalent to considering the modelled impulse response as a realisation of a Gaussian process with zero mean and covariance $P$ \cite{Pil10}.

In this Bayesian framework, it is possible to include the assumed system properties by means of a clever parametrisation of $P$. Since it is reasonable to assume that the true impulse response is smooth, and exponentially decaying to zero (for stable systems), one can parametrise $P$ as follows (DC kernel):
\begin{equation}
\label{DC}
P_{DC}(i,j)=c\rho^{|i-j|}\alpha^{(i+j)/2}.
\end{equation}
The special case for which $\rho=\sqrt{\alpha}$ results in the so-called TC kernel:
\begin{equation}
\label{TC}
P_{TC}(i,j)=c\min{(\alpha^i,\alpha^j)}.
\end{equation}
Here $c\geq0$, $|\rho|\leq 1$ and $0\leq\alpha\leq1$ are hyperparameters that need to be tuned based on the available data, e.g.\ by marginal likelihood maximisation (Empirical Bayes method). See \cite{Chen12} and \cite{Chen13} for more details on this issue.
%
%
%

\subsection{Filter interpretation of the cost function}
\label{secfil}

Alternatively, one can include the available prior knowledge about the system directly in the regularisation matrix $R$. In this way, instead of defining the problem based on the kernel $P$, one can focus directly on the cost function.

The approach proposed here starts from the following decomposition of the $n \times n$ matrix $R$:
\begin{equation}
\label{RdecF}
R=\lambda F^T F,
\end{equation}
where $F$ is a $n \times n$ matrix, and $\lambda \in \mathbb{R}$ is a global scaling factor for the regularisation matrix.

The cost function is then reformulated as follows:
\begin{align}
\label{filtercost}
\|Y-\Phi\theta\|^2+\theta^TR\theta &=\|Y-\Phi\theta\|^2+\lambda\theta^TF^T F\theta \\ &=\|Y-\Phi\theta\|^2+\lambda\|F\theta\|^2.
\end{align}

In the above equation, $F$ can be seen as a prefiltering operator on the coefficients of the impulse response, before they enter the cost function and are penalised through the regularisation term. 
This means that the regularisation filter matrix $F$ should be defined in such a way that it incorporates the system properties one needs to penalise in order to obtain the desired model. 

Next section provides a first intuitive understanding of how this should be achieved. A more detailed discussion about the implementation of the filter interpretation ideas to design a novel regularised FIR modelling approach is presented in Section~\ref{sec5}.

\section{Understanding the filtering approach for regularised FIR modelling}
\label{sec4}


\subsection{Smoothness}

\subsubsection{An illustrative example: the `random walk' kernel}

Let us start from a simple introductory example that will prove useful in explaining the basic principle of the filtering approach.

Consider the case in which each of the parameters $\theta$ can be described by a random walk:
\begin{equation}
\theta_k=\sum_{l=1}^{k} e_l
\end{equation}
where one has $E(e_l)=0$ and $E(e_l^2)=\sigma_e^2$.

By using the well-known properties of random walks, one can compute the covariance matrix of $\theta$ as follows:
\begin{equation}
P_{RW}(i,j)=\text{Cov}_{ij}=E(\theta_i,\theta_j)=\min(i,j) \cdot \sigma_e^2.
\end{equation}
If one computes the inverse of such a covariance matrix, and using the definition of the regularisation matrix in (\ref{Rinv}), the following expression is obtained:
\begin{equation}
R_{RW}=\sigma^2(P_{RW})^{-1}=\frac{\sigma^2}{\sigma_e^2}
  \begin{pmatrix}
   2 & -1 & 0 & \cdots & 0 \\
   -1 & 2 & -1 & \ddots & \vdots \\ 
   0 & \ddots & \ddots & \ddots & 0 \\ 
\vdots & \ddots & -1 & 2 & -1\\
 0 & \cdots & 0 & -1 & 1
  \end{pmatrix}.
\end{equation}
This means that the second term in the cost function (\ref{filtercost}) can be written as:
\begin{equation}
\theta^TR_{RW}\theta=\sum_{k=1}^{n} (\theta_k-\theta_{k-1})^2
\end{equation}
where $\theta_0$ has been set equal to $0$, and $\sigma^2/\sigma_e^2=1$, for simplicity.

The main message here is that what is penalised in the cost function is actually the squared difference of subsequent values of $\theta$. This is one way of imposing smoothness on the parameters, since large changes between two adjacent impulse response coefficients are penalised through $R_{RW}$. Of course at this level nothing is required about the decay of the impulse response, this will be taken into account later on in this section.

Bearing the interpretation introduced in Section~\ref{secfil} in mind, $R_{RW}$ is factorised to obtain the filtering matrix $F_{RW}$:
\begin{equation}
F_{RW}(i,j) =
  \begin{cases}
     1       & \quad \text{for } i=j \\
     -1  & \quad \text{for } i=j+1 \\
     0 & \quad \text{otherwise. }\\
   \end{cases}
\end{equation}

Note that here and in the remainder of this section the scaling factor $\lambda$ is set equal to $1$ to simplify the notation.

\begin{figure}[t]
\centering
\subfigure[]{
\includegraphics[width=3cm]{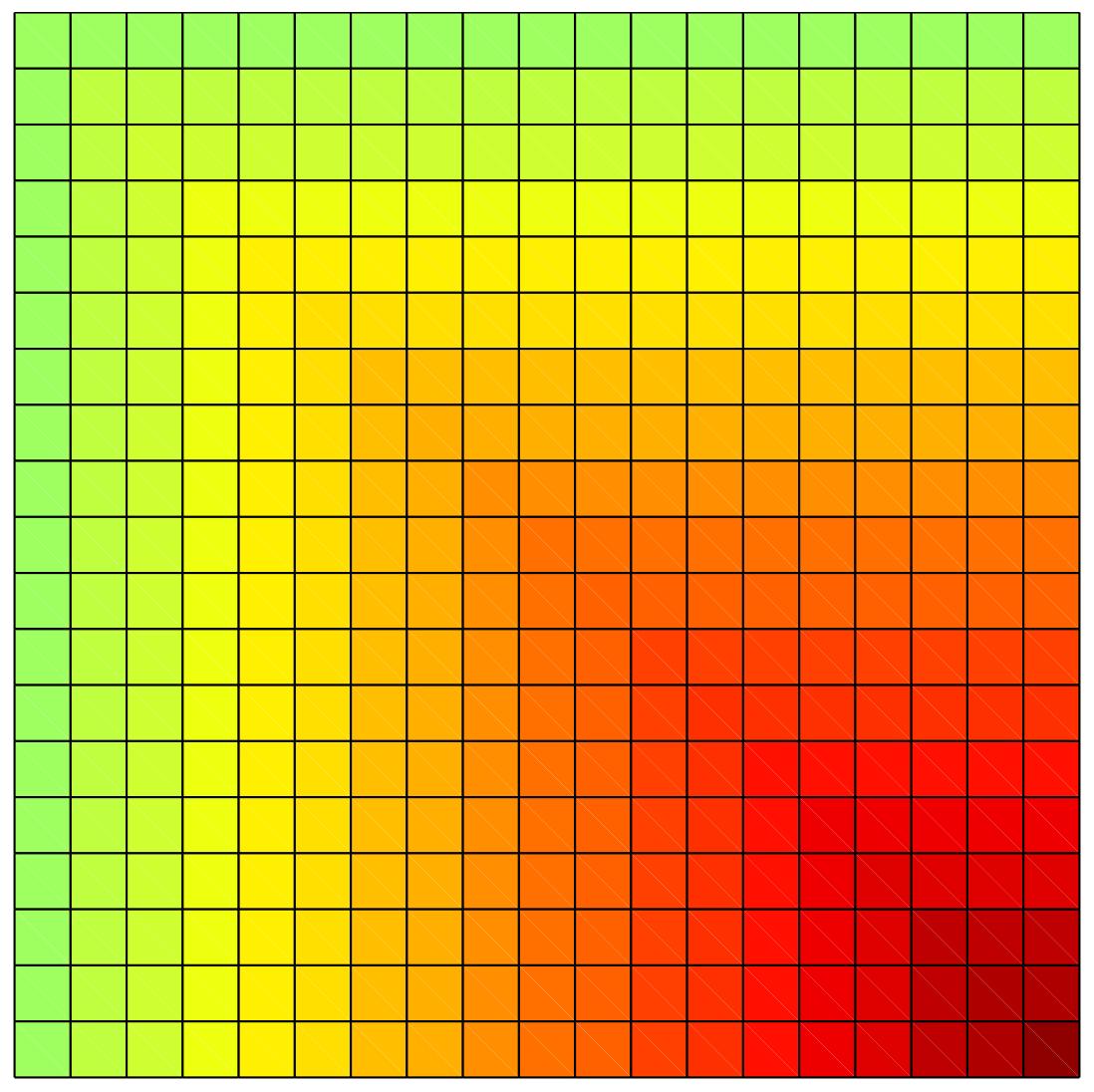}
\label{figrw}
}
\subfigure[]{
\includegraphics[width=3cm]{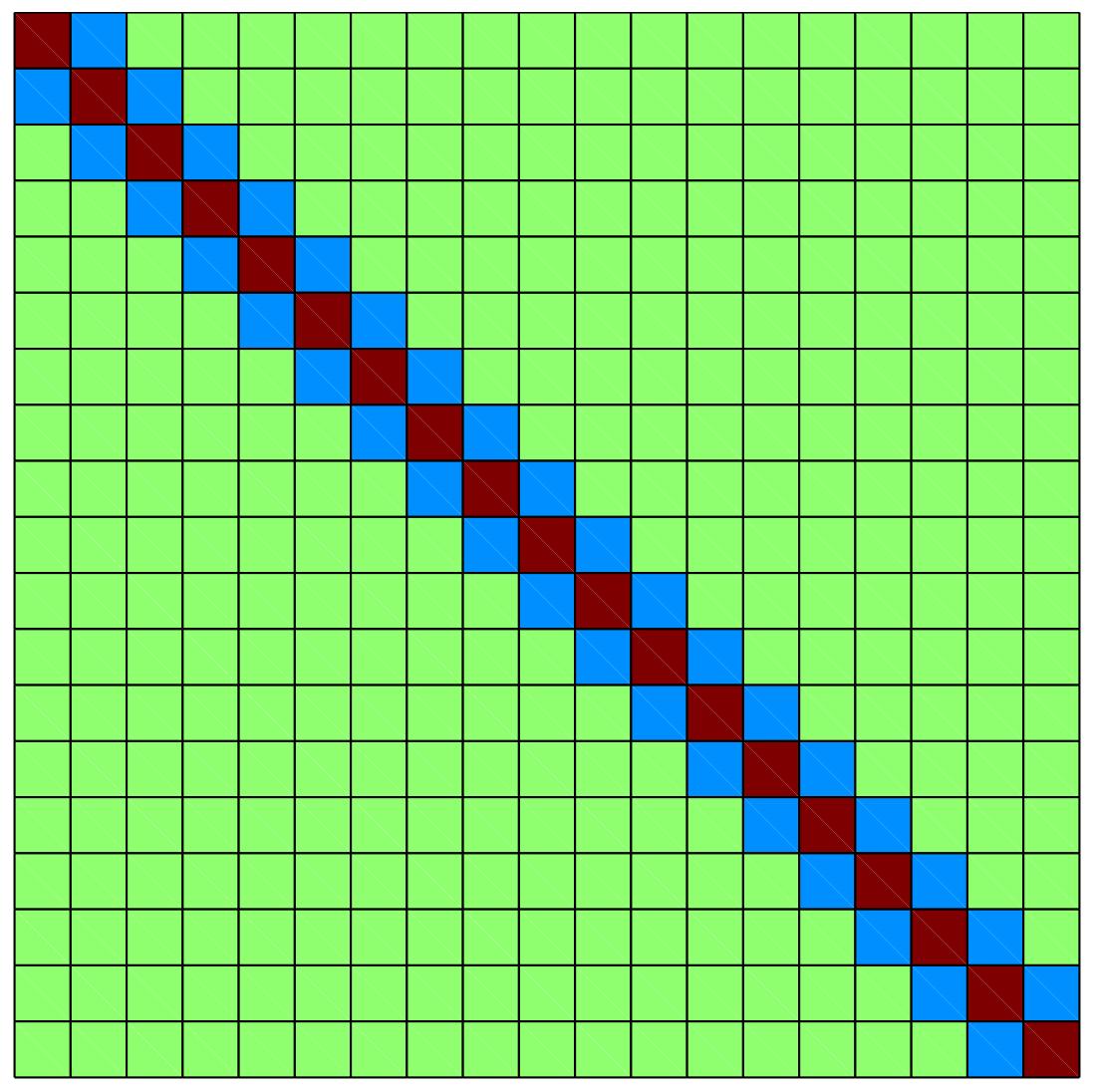}
\label{figrwi}
} 
\subfigure[]{
\includegraphics[width=3cm]{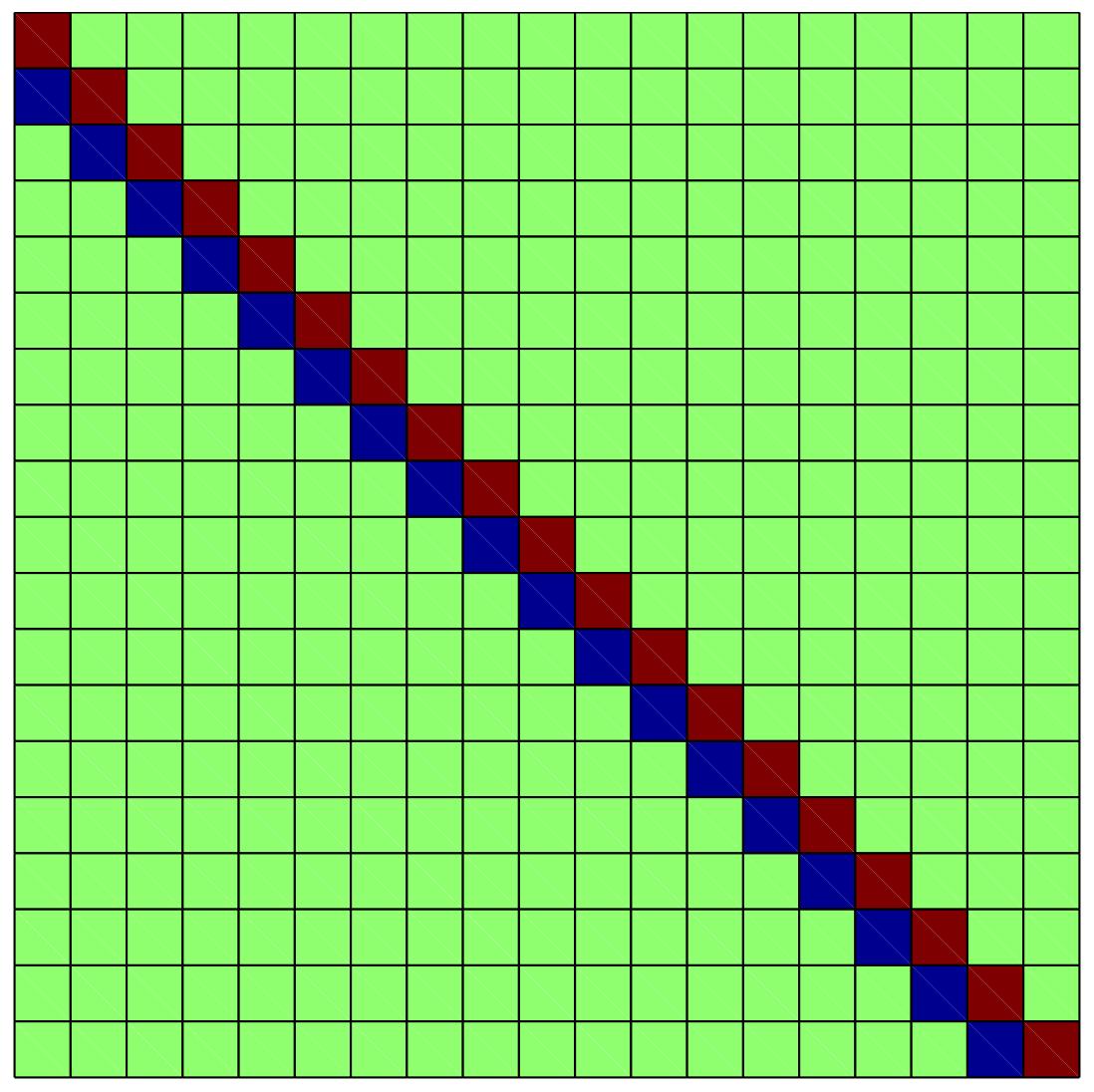}
\label{figrwf}
} 
\subfigure[]{
\includegraphics[width=7.5cm]{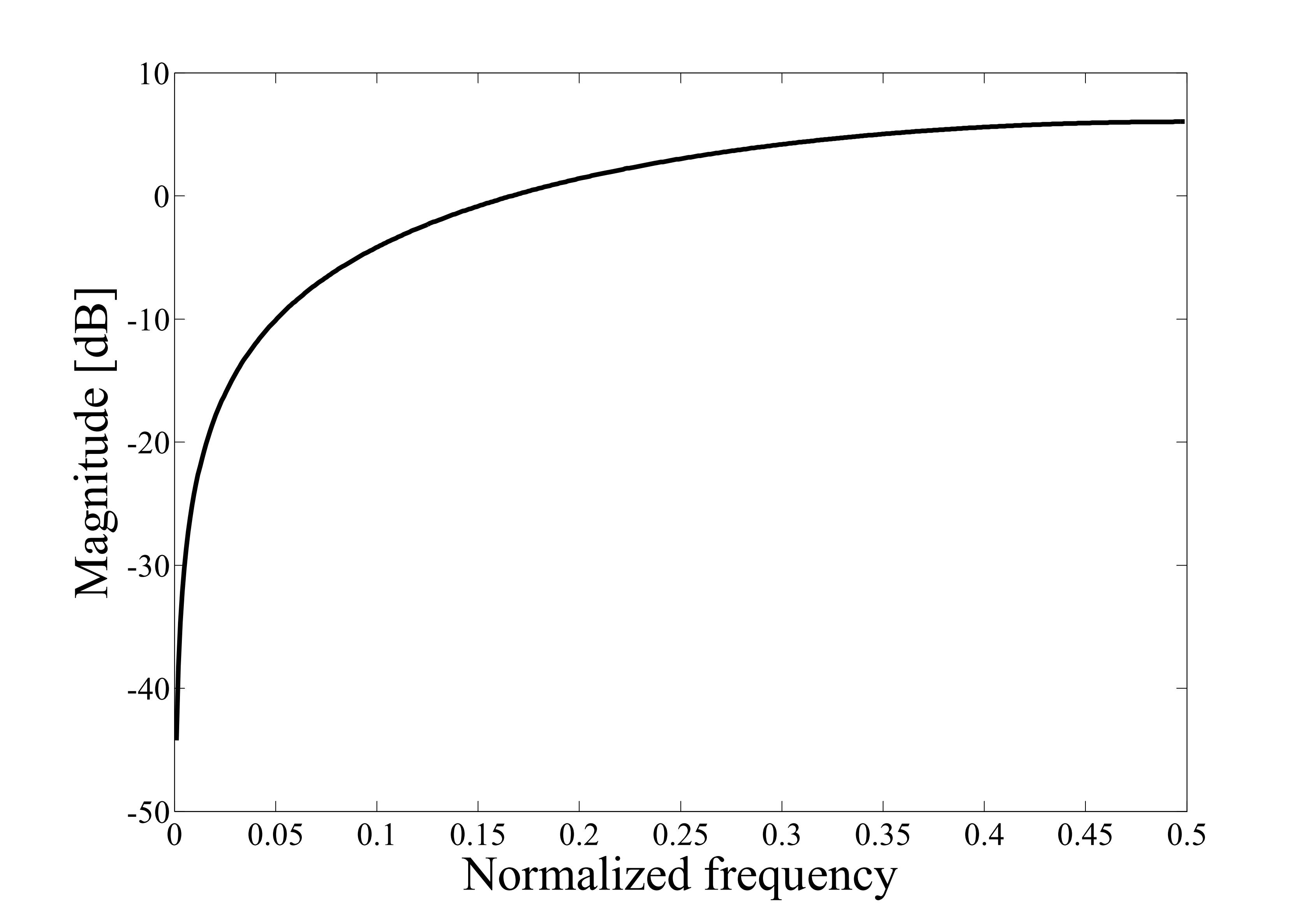}
\label{figfiltrw}
}
\caption{Random walk kernel: (a) covariance matrix of the parameters $P_{RW}$; (b) regularisation matrix $R_{RW}$; (c) filter matrix $F_{RW}$, for $n=20$, and for specific values $\sigma^2=\sigma_e^2=1$. The colour map should be read as follows; green: zero values, (darker) red: (larger) positive values, (darker) blue: (larger) negative values. (d) Magnitude (in dB) of the frequency response of the rows of the filter matrix $F_{RW}$.}
\label{figRW}
\end{figure}

Figure~\ref{figRW} shows the structure of matrices $P_{RW}$, $R_{RW}$ and $F_{RW}$. Note the tridiagonal structure of $R_{RW}$, and the bidiagonal structure of $F_{RW}$.

In general, one can decide to compute the Cholesky decomposition to obtain $F$ from $R$ ($R=LL^T$, with $L$ a lower triangular matrix, and $F=L^T$) \cite{golub96}. In this example, however, a rotated version is preferred ($R=L^TL$, with $F=L$), to show more clearly the nature of the filtering matrix. 

Each row of $F_{RW}$ contains the filter coefficients related to a pair of subsequent $\theta$ values (since $F_{RW}$ is bidiagonal). Therefore, it seems natural to study how the filtering operation works on $\theta$ by having a look at the frequency response of the rows of $F_{RW}$, plotted in Figure~\ref{figfiltrw}. The high-pass nature of the filtering is evident, which indicates that only the high frequency components in the modelled impulse response enter the cost function and are thus penalised (i.e.\ the smoothness property is imposed). 

%
%

This simple example illustrates how known concepts can be reinterpreted using the filtering ideas, and gives already an idea of how specific assumptions about the system can be encoded at the cost function level.

%
%


\subsubsection{The correlation kernel}

Let us now have a look at another way of imposing the smoothness property on the impulse response, which brings us closer to the definition of the kernels that are available in the literature.

Consider a `correlation' kernel expressed as:
\begin{equation}
P_{Corr}(i,j)=c\rho^{|i-j|}
\end{equation}
where $c$ is a positive constant, and $\rho$ is a value ($|\rho|\leq 1$) that can be tuned to specify how strong the correlation between variables is. Since $P_{Corr}$ represents the covariance matrix of the parameters $\theta$, here smoothness is imposed in the sense that coefficients of the impulse response closer to each other are more strongly correlated (the distance between $i$ and $j$ is smaller) than coefficients that are further away from each other (larger distance between $i$ and $j$).

The analytical expression for the regularisation matrix $R_{Corr}$ can be computed as:
\begin{equation}
R_{Corr}=\sigma^2(P_{Corr})^{-1}
=\frac{\sigma^2}{c}
  \begin{pmatrix}
   \frac{1}{1-\rho^2} & -\frac{\rho}{1-\rho^2} & 0 & \cdots & 0 \\
   -\frac{\rho}{1-\rho^2} & \frac{1+\rho^2}{1-\rho^2} & -\frac{\rho}{1-\rho^2} & \ddots & \vdots \\ 
   0 & \ddots & \ddots & \ddots & 0 \\ 
\vdots & \ddots & -\frac{\rho}{1-\rho^2} & \frac{1+\rho^2}{1-\rho^2} & -\frac{\rho}{1-\rho^2}\\
 0 & \cdots & 0 & -\frac{\rho}{1-\rho^2} & \frac{1}{1-\rho^2}
  \end{pmatrix}.
\end{equation}

Since the regularisation matrix has a tridiagonal structure, its Cholesky decomposition can easily be computed in closed form, as explained e.g.\ in \cite{Loan00}:
\begin{equation}
F_{Corr}(i,j) =
  \begin{cases}
     \sqrt{\frac{1}{1-\rho^2}}  & \quad \text{for } i=j, j<n \\
     1  & \quad \text{for } i=j=n \\
     -\sqrt{\frac{\rho^2}{1-\rho^2}}  & \quad \text{for } i=j-1 \\
     0 & \quad \text{otherwise. }\\
   \end{cases}
\end{equation}
Note that in the expression for $F_{Corr}$ the values of $c$ and $\sigma^2$ are set equal to 1 to simplify the notation.

\begin{figure}[t]
\centering
\subfigure[]{
\includegraphics[width=3cm]{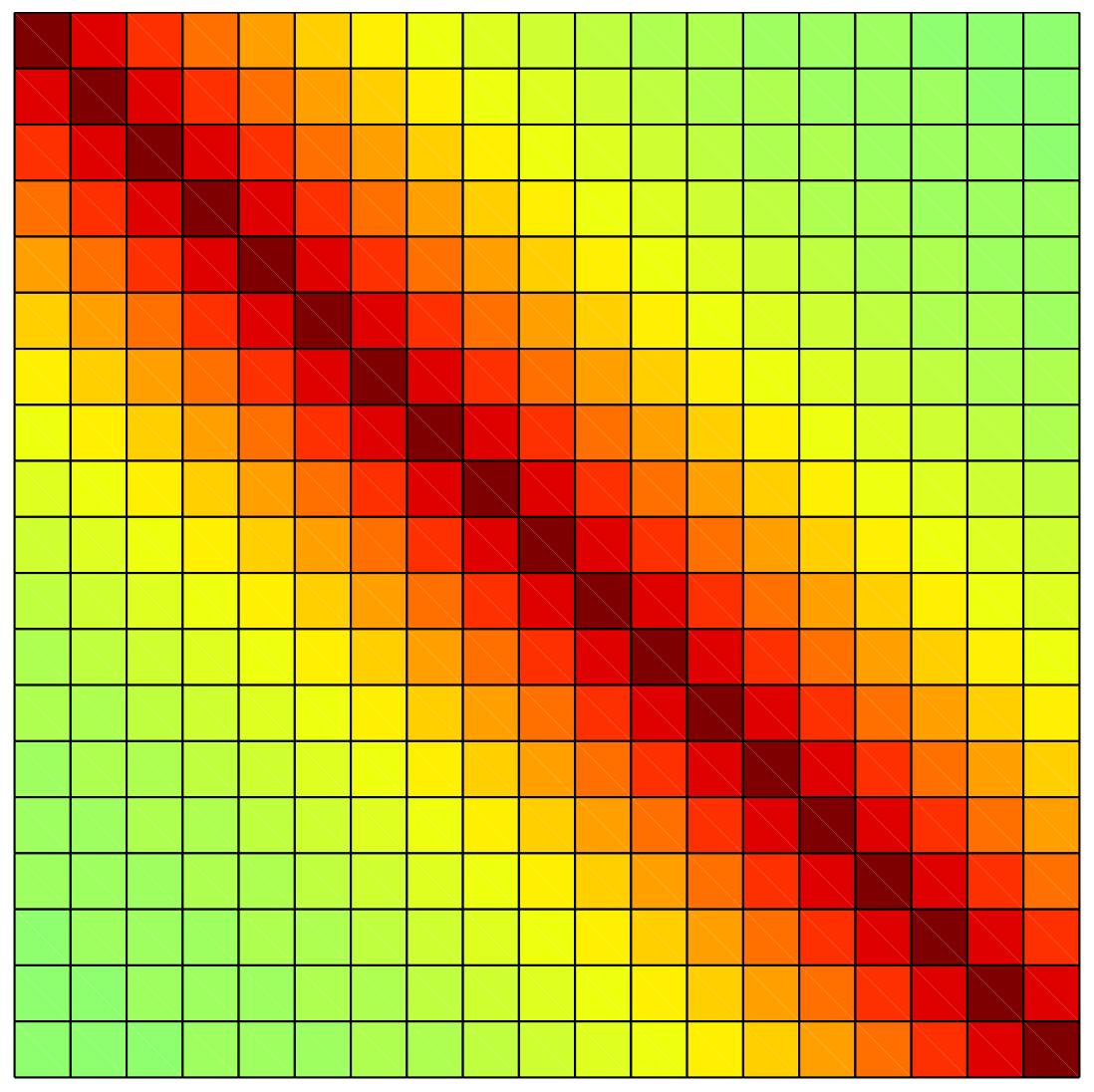}
\label{figcorr}
}
\subfigure[]{
\includegraphics[width=3cm]{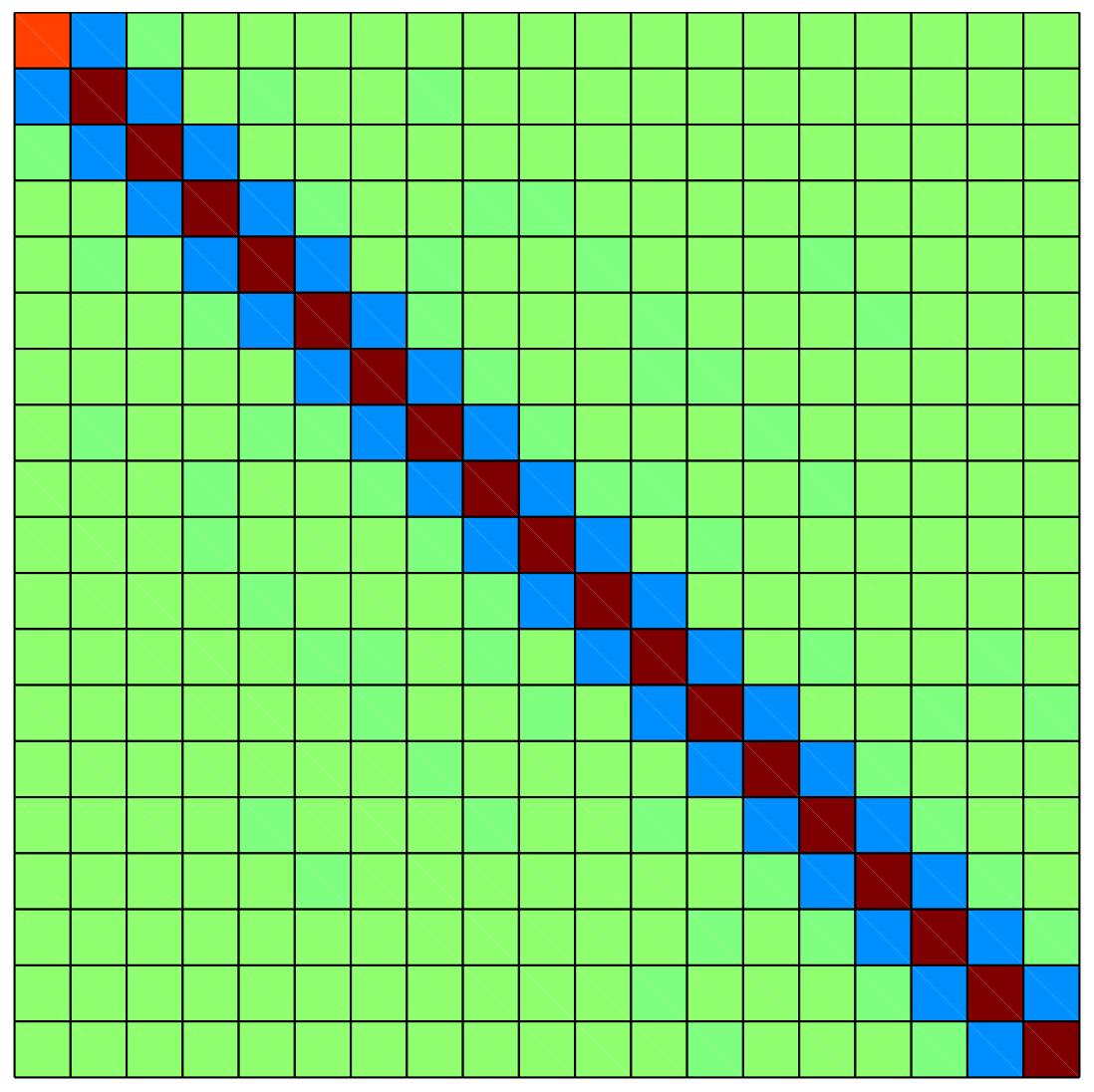}
\label{figcorri}
} 
\subfigure[]{
\includegraphics[width=3cm]{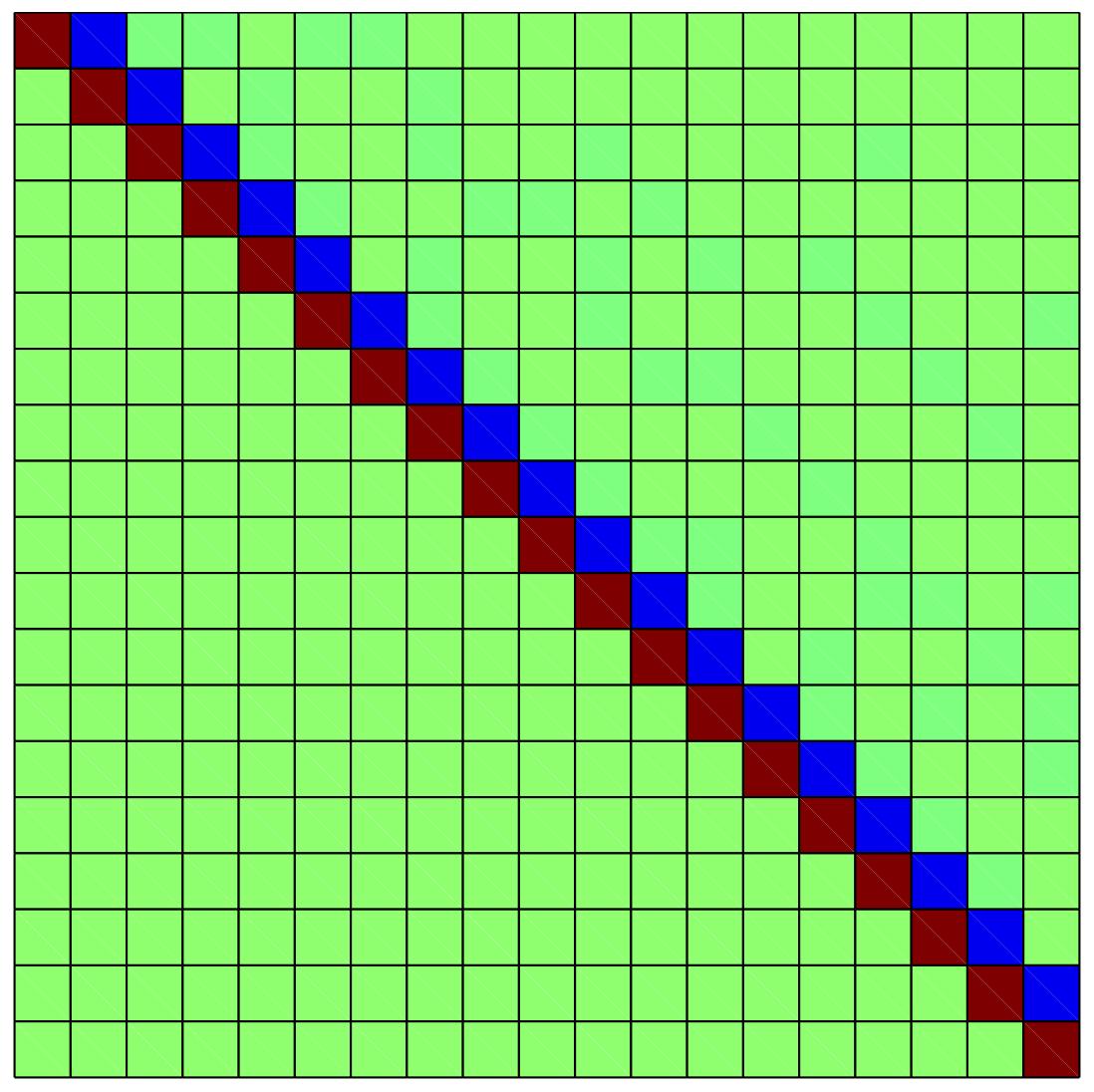}
\label{figcorrf}
} 
\subfigure[]{
\includegraphics[width=7.5cm]{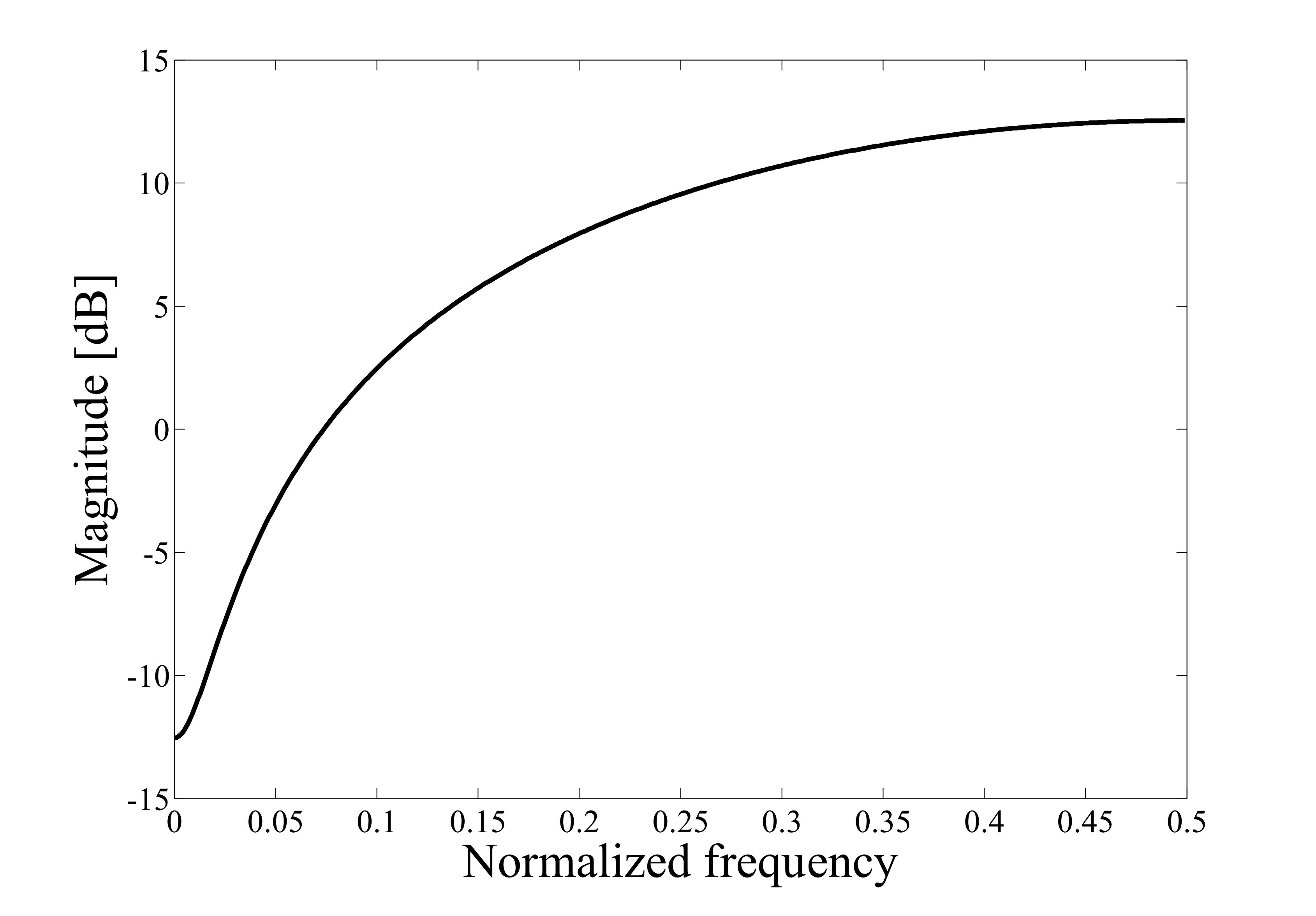}
\label{figfiltcorr}
} 
\caption{Correlation kernel: (a) covariance matrix of the parameters $P_{Corr}$; (b) regularisation matrix $R_{Corr}$; (c) filter matrix $F_{Corr}$, for $n=20$, and for specific values $c=1$, $\sigma^2=1$ and $\rho=0.8$. The colour map should be read as follows; green: zero values, (darker) red: (larger) positive values, (darker) blue: (larger) negative values. (d) Magnitude (in dB) of the frequency response of the rows of the filter matrix $F_{Corr}$.}
\label{figCorr}
\end{figure}

The structure of matrices $P_{Corr}$, $R_{Corr}$ and $F_{Corr}$ is shown in Figure~\ref{figCorr}.

To get an idea of how the filtering matrix acts on the parameters $\theta$, the frequency response of the rows of $F_{Corr}$ is presented in Figure~\ref{figfiltcorr}.

%

As expected, this is, similarly to the random walk example, a high-pass filter, which is the translation of the smoothness property in the filtering interpretation suggested in this paper.

\subsection{Decay}

For the moment being, let us put the concept of smoothness aside, and analyse a different assumption that is typically made about stable impulse responses, namely the idea that they should exponentially decay to zero. 

This can be expressed very simply by the following covariance matrix of the parameters:
\begin{equation}
P_{Dec}(i,j)=
  \begin{cases}
     \alpha^i  & \quad \text{for } i=j\\
     0 & \quad \text{otherwise }\\
   \end{cases}
\end{equation}
where $\alpha$ is a value ($0\leq\alpha\leq1$) that can be tuned to specify how fast the coefficients $\theta$ decay to zero.

In this case, both the regularisation matrix and the filter matrix are obtained in a straightforward way:
\begin{equation}
R_{Dec}(i,j)= \sigma^2
  \begin{cases}
     \alpha^{-i}  & \quad \text{for } i=j\\
     0 & \quad \text{otherwise }\\
   \end{cases}
\end{equation}

\begin{equation}
F_{Dec}(i,j)=
  \begin{cases}
     \alpha^{-i/2}  & \quad \text{for } i=j\\
     0 & \quad \text{otherwise. }\\
   \end{cases}
\end{equation}
Again, in the expression for $F_{Dec}$ the value of $\sigma^2$ is set equal to 1 to simplify the notation.

\begin{figure}[t]
\centering
\subfigure[]{
\includegraphics[width=3cm]{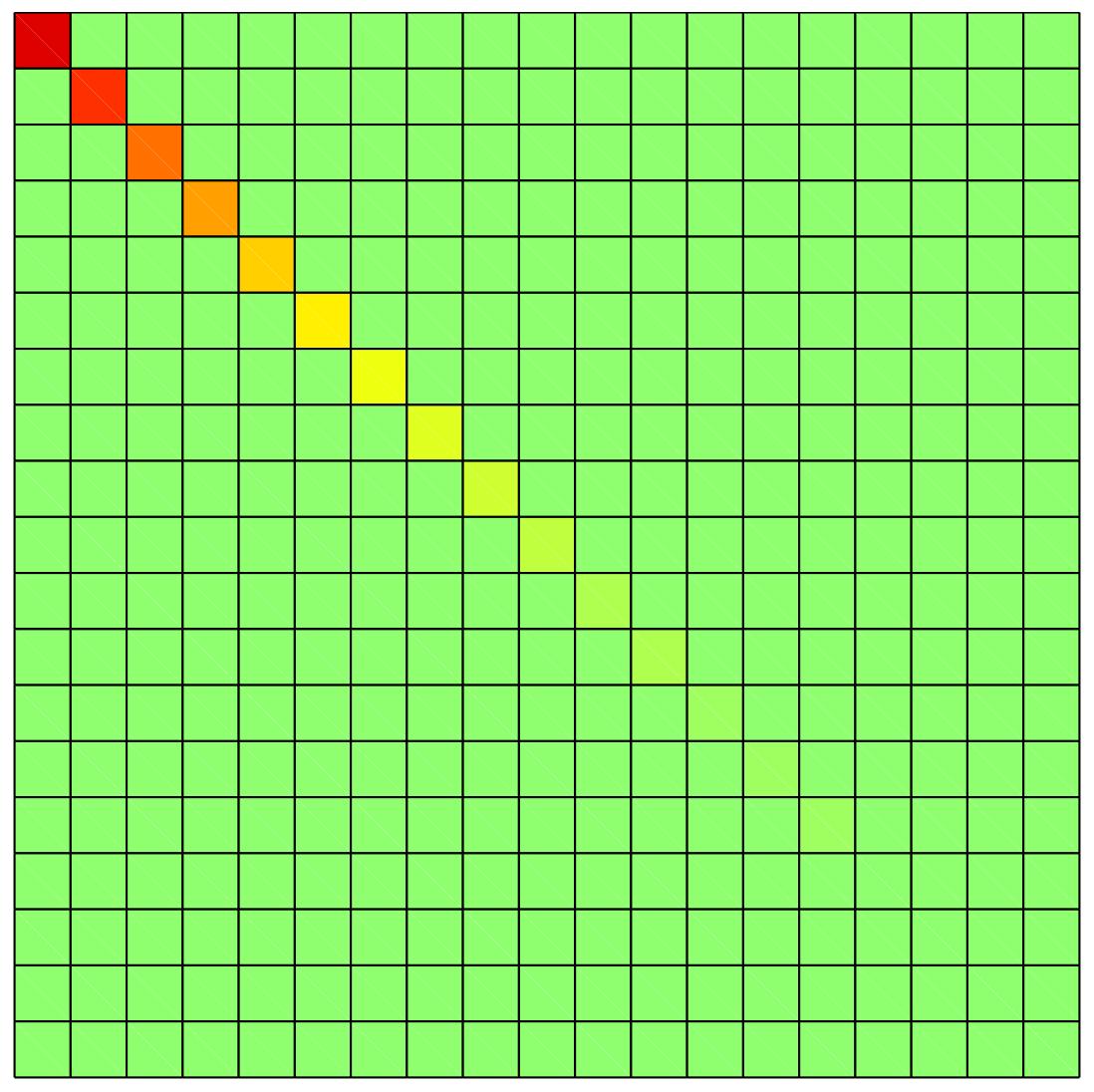}
\label{figdec}
}
\subfigure[]{
\includegraphics[width=3cm]{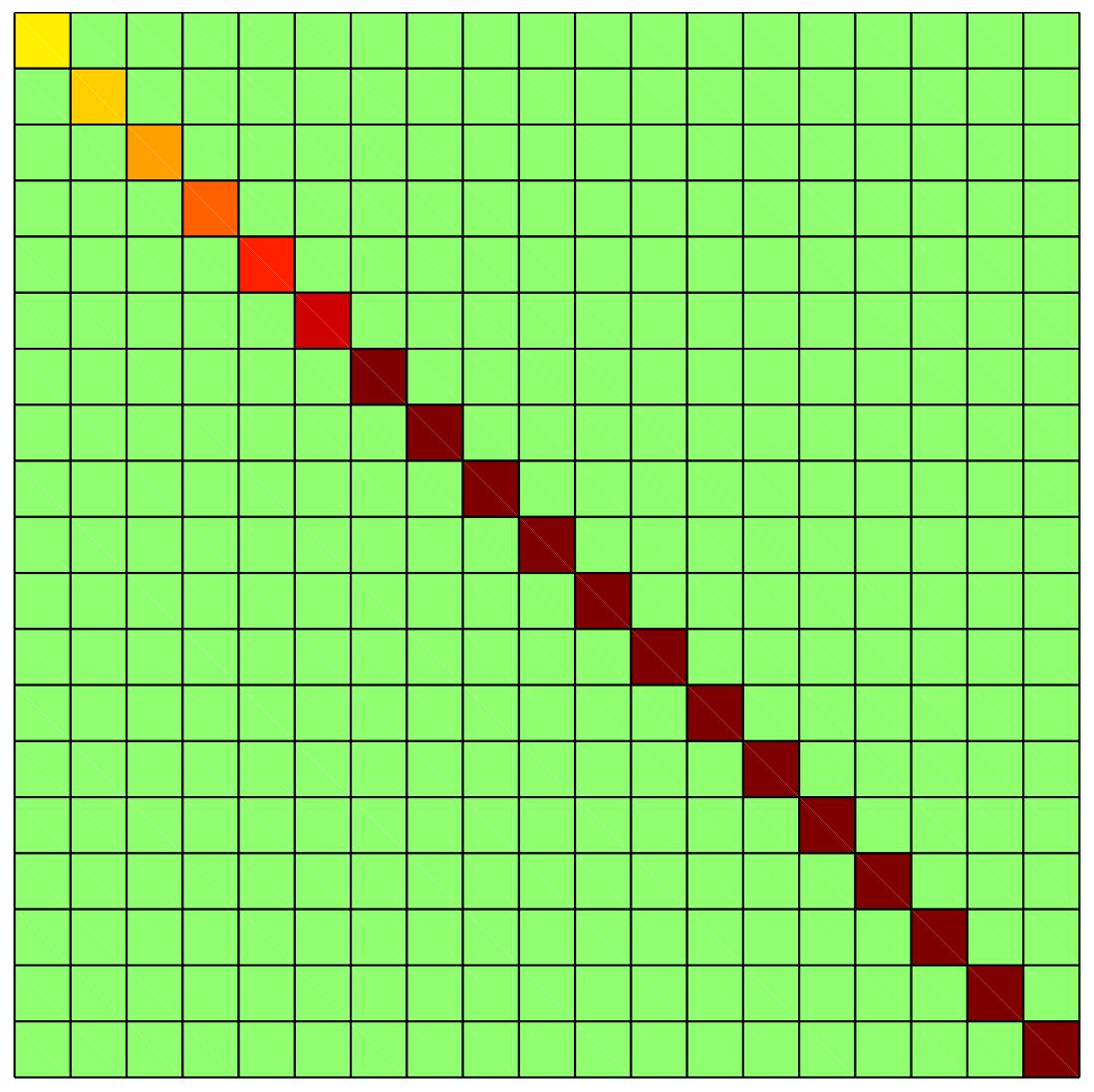}
\label{figdeci}
} 
\subfigure[]{
\includegraphics[width=3cm]{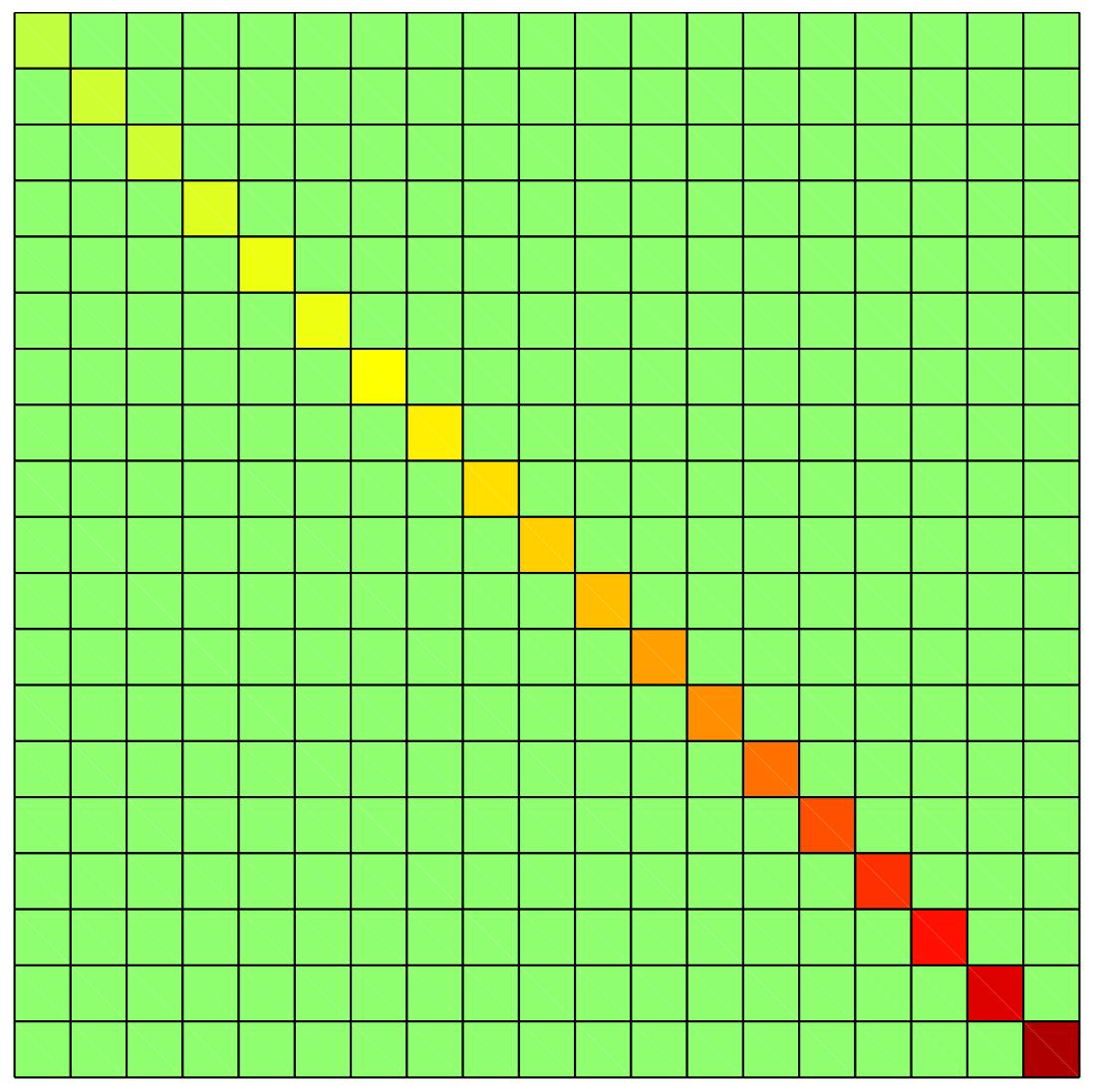}
\label{figdecf}
} 
\subfigure[]{
\includegraphics[width=7.5cm]{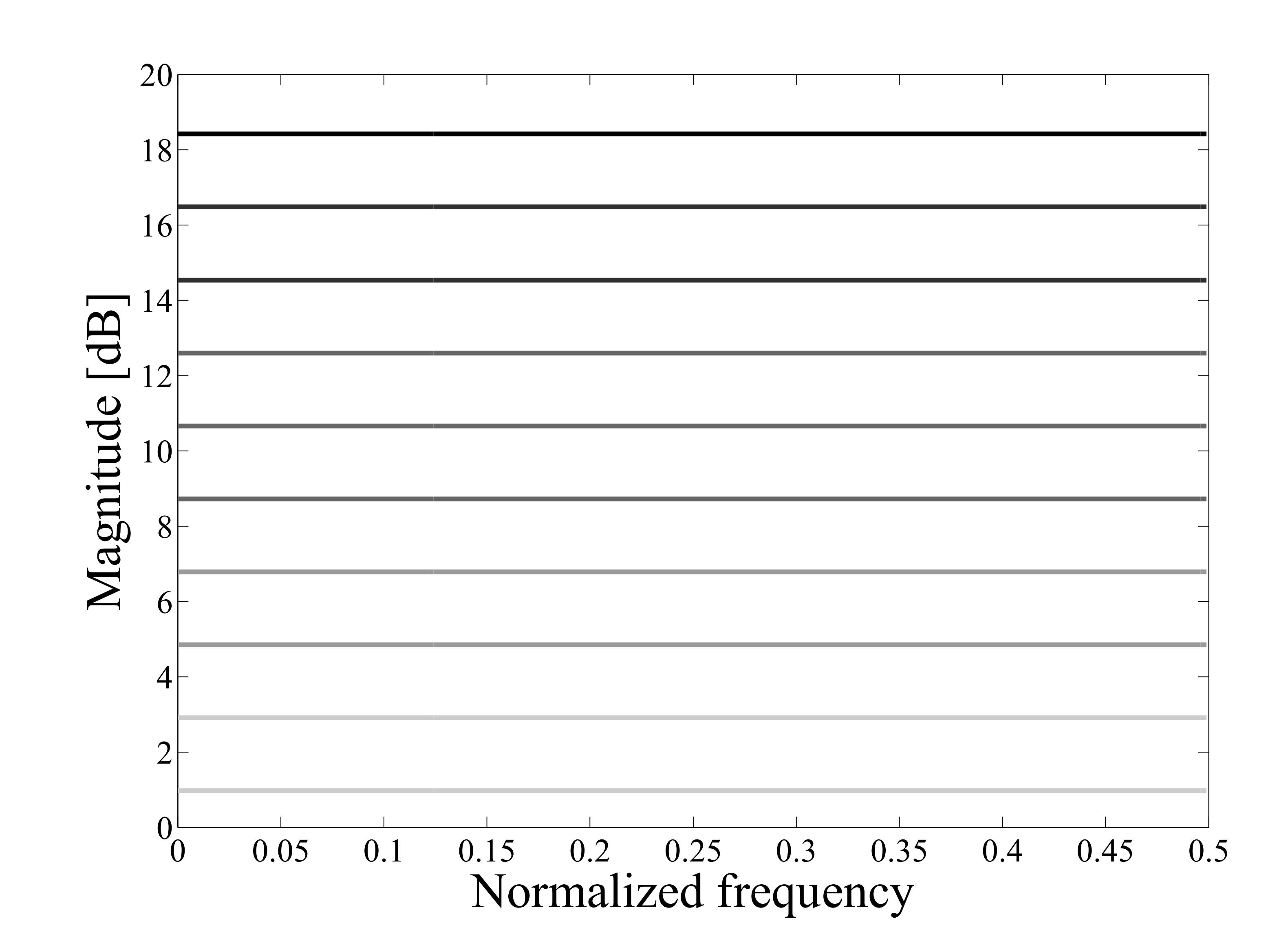}
\label{figfiltdec}
} 
\caption{Decay kernel: (a) covariance matrix of the parameters $P_{Dec}$; (b) regularisation matrix $R_{Dec}$; (c) filter matrix $F_{Dec}$, for $n=20$, and for specific values $\sigma^2=1$ and $\alpha=0.8$. The colour map should be read as follows; green: zero values, (darker) red: (larger) positive values, (darker) blue: (larger) negative values. (d) Magnitude (in dB) of the frequency response of the rows of the filter matrix $F_{Dec}$. To simplify the plot, only the odd-numbered rows are considered. A darker line is used to indicate rows with increasing number.}
\label{figDec}
\end{figure}

The structure of matrices $P_{Dec}$, $R_{Dec}$ and $F_{Dec}$ is shown in Figure~\ref{figDec}, together with the frequency response of the different rows.

Since each row of the filtering matrix contains only a coefficient ($\alpha^{-i/2}$, which increases for higher values of $i$), each frequency response is a constant gain factor, with higher gain for rows with increasing number, i.e.\ higher gain towards the tail of the impulse response (last coefficients in $\theta$). This means that the coefficients at the tail of the impulse response are more heavily penalised than the coefficients at the beginning of the impulse response (i.e.\ the exponential decay property is imposed).

\subsection{Smoothness and decay}

Having studied the two different properties (smoothness and decay) separately in the previous parts of this section, let us now combine the knowledge gathered so far, to analyse the widely-used DC and TC kernels \cite{Chen12} using the proposed filtering interpretation.

Let us consider the DC kernel in (\ref{DC}).

Given (\ref{Rinv}) and (\ref{DC}), the regularisation matrix $R_{DC}$ can be analytically computed as:
\begin{equation}
R_{DC}(i,j)
=\frac{\sigma^2}{c} \frac{a_{ij}}{\alpha^{(i+j)/2}(1-\rho^2)}
\end{equation}
where
\begin{equation} a_{ij}=
\begin{cases}
     1+\rho^2  & \quad \text{for } i=j, 1<j<n \\
     1  & \quad \text{for } i=j=1 \text{ and } i=j=n \\
     -\rho  & \quad \text{for } \left|i-j\right|=1 \\
     0 & \quad \text{otherwise. }\\
   \end{cases}
\end{equation}

The filter matrix is computed as the Cholesky decomposition of $R_{DC}$:
\begin{equation}
F_{DC}(i,j) =
  \begin{cases}
     \sqrt{\frac{1}{\alpha^i(1-\rho^2)}}  & \quad \text{for } i=j, j<n \\
     \sqrt{\frac{1}{\alpha^i}}  & \quad \text{for } i=j=n \\
     -\sqrt{\frac{\rho^2}{\alpha^{i+1}(1-\rho^2)}}  & \quad \text{for } i=j-1 \\
     0 & \quad \text{otherwise. }\\
   \end{cases}
\label{DCfiltermat}
\end{equation}
Again, $c$ and $\sigma^2$ are set equal to 1 to simplify the notation.

As already mentioned in Section~\ref{secker}, if one chooses $\rho=\sqrt{\alpha}$, the TC kernel in (\ref{TC}) is obtained.

The corresponding regularisation matrix $R_{TC}$ therefore becomes:
\begin{equation}
R_{TC}(i,j)
=\frac{\sigma^2}{c} \frac{a_{ij}}{\alpha^{(i+j)/2}(1-\alpha)}
\end{equation}
where
\begin{equation} a_{ij}=
\begin{cases}
     1+\alpha  & \quad \text{for } i=j, 1<j<n \\
     1  & \quad \text{for } i=j=1 \text{ and } i=j=n \\
     -\sqrt{\alpha}  & \quad \text{for } \left|i-j\right|=1 \\
     0 & \quad \text{otherwise }\\
   \end{cases}
\end{equation}
and its Cholesky decomposition is:
\begin{equation}
F_{TC}(i,j) =
  \begin{cases}
     \sqrt{\frac{1}{\alpha^i(1-\alpha)}}  & \quad \text{for } i=j, j<n \\
     \sqrt{\frac{1}{\alpha^i}}  & \quad \text{for } i=j=n \\
     -\sqrt{\frac{1}{\alpha^i(1-\alpha)}}  & \quad \text{for } i=j-1 \\
     0 & \quad \text{otherwise }\\
   \end{cases}
\label{TCfiltermat}
\end{equation}
with $c$ and $\sigma^2$ equal to 1.

Note that analogous analytical expressions for the inverse of $P_{TC}$, here denoted $R_{TC}$, and for the factorisation of $R_{TC}$ were also independently derived in \cite{Chen16} and \cite{Carli16}, where maximum entropy properties of TC and DC kernels are discussed and such a factorisation is used to reduce the computational complexity of the estimation algorithm.

\begin{figure}[t]
\centering
\subfigure[]{
\includegraphics[width=3cm]{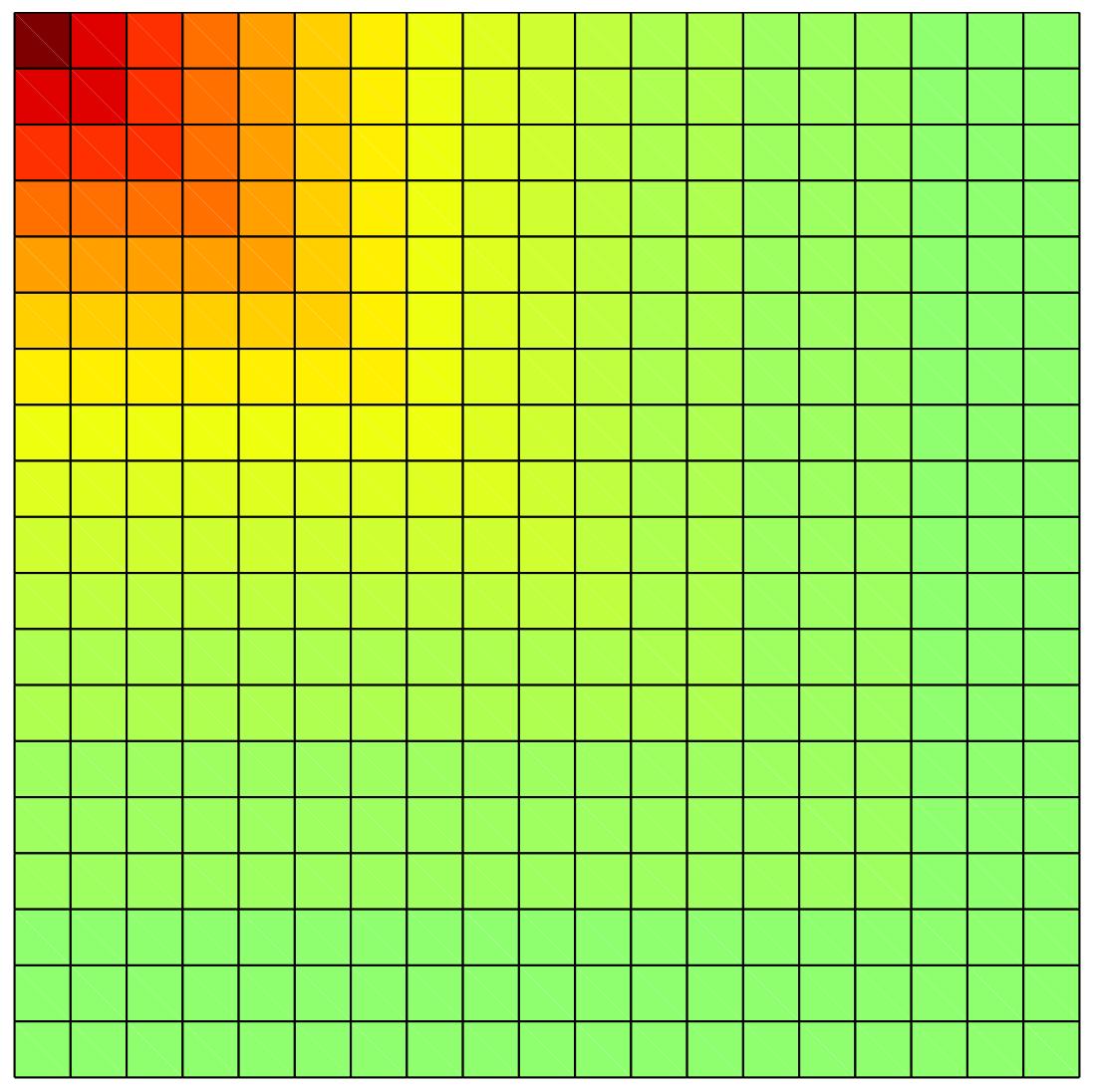}
\label{figtc}
}
\subfigure[]{
\includegraphics[width=3cm]{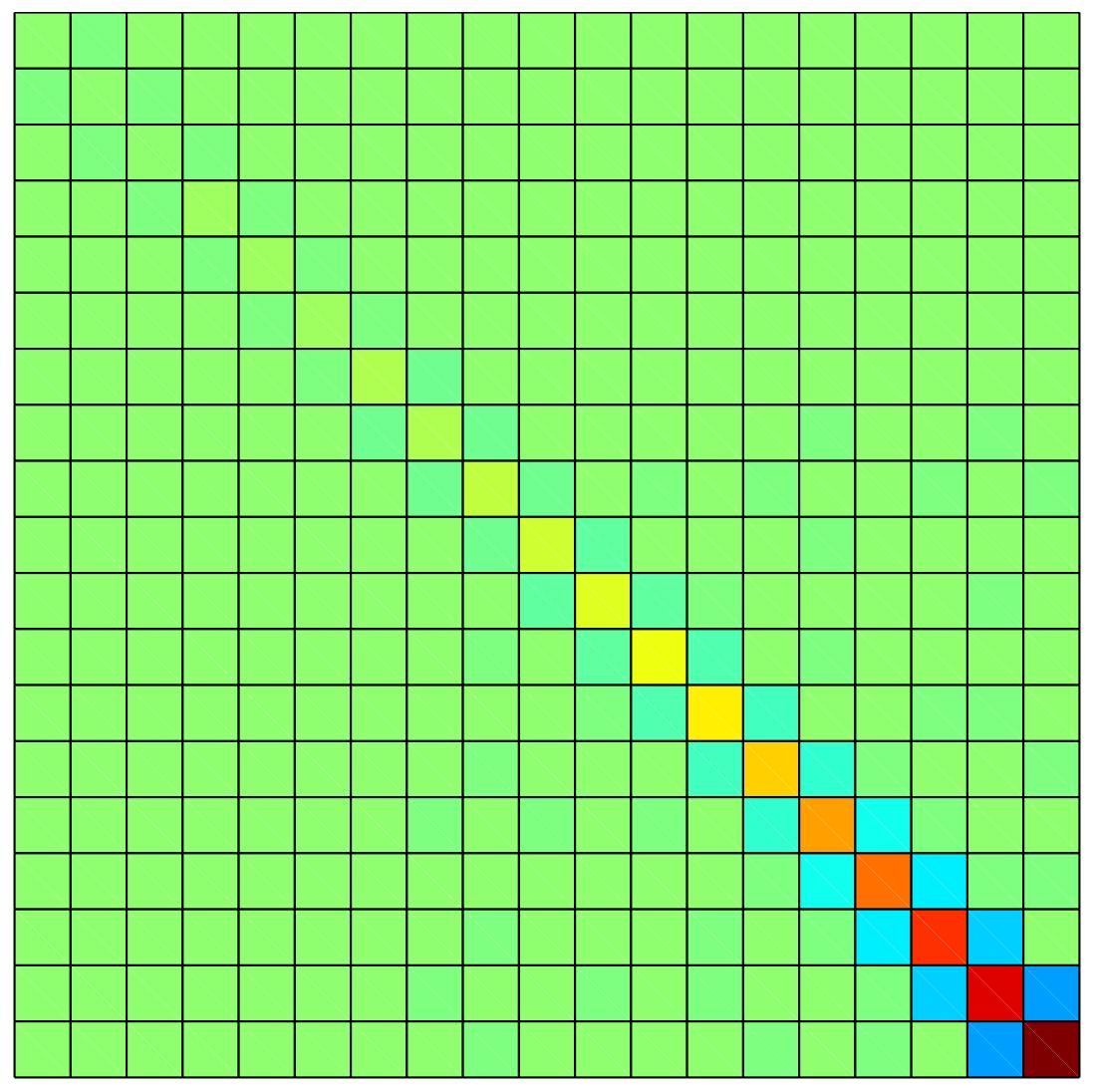}
\label{figtci}
} 
\subfigure[]{
\includegraphics[width=3cm]{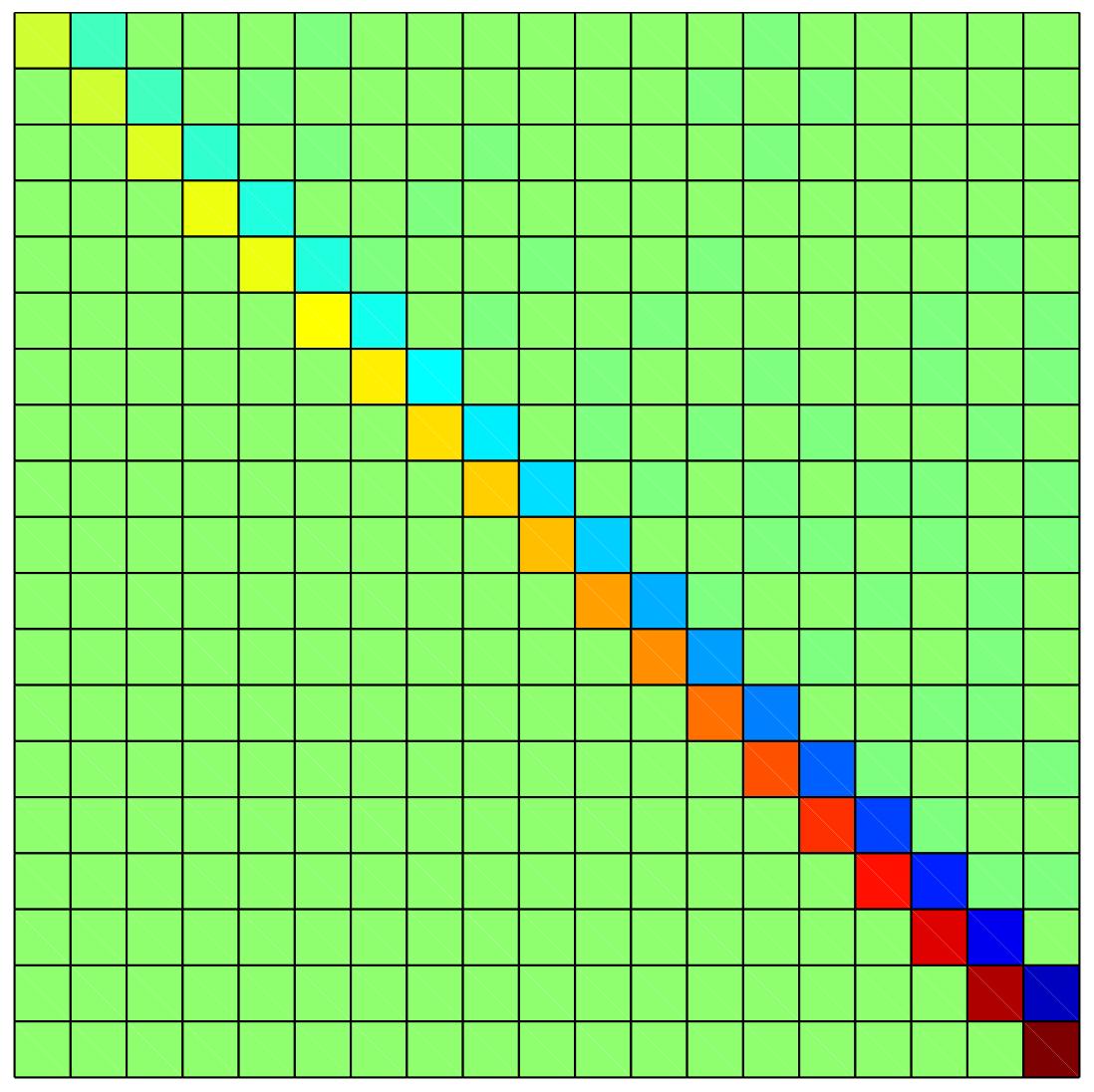}
\label{figtcf}
} 
\subfigure[]{
\includegraphics[width=7.5cm]{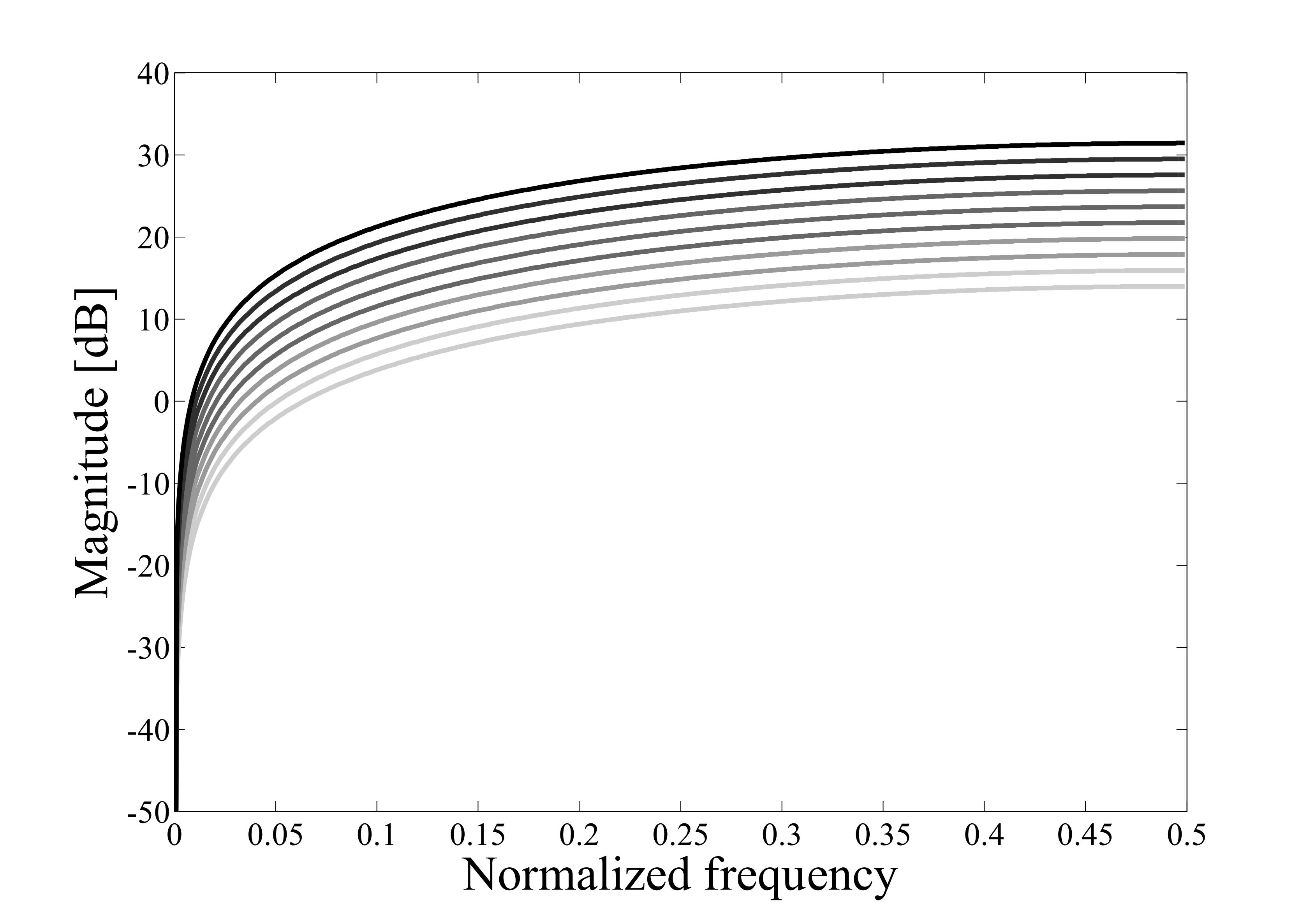}
\label{figfilttc}
} 
\caption{TC kernel: (a) covariance matrix of the parameters $P_{TC}$; (b) regularisation matrix $R_{TC}$; (c) filter matrix $F_{TC}$, for $n=20$, and for specific values $c=1$, $\sigma^2=1$ and $\alpha=0.8$. The colour map should be read as follows; green: zero values, (darker) red: (larger) positive values, (darker) blue: (larger) negative values. (d) Magnitude (in dB) of the frequency response of the rows of the filter matrix $F_{TC}$. To simplify the plot, only the odd-numbered rows are considered. A darker line is used to indicate rows with increasing number.}
\label{figTc}
\end{figure}

The structure of matrices $P_{TC}$, $R_{TC}$ and $F_{TC}$ is shown in Figure~\ref{figTc}.

%

For the sake of simplicity here we will address only the TC kernel example, but all the following considerations can of course be generalised to the DC kernel case. 

Figure~\ref{figfilttc} shows the frequency response of each row in $F_{TC}$.

One can observe the high-pass nature of the filtering and a higher gain of the response for rows with increasing number. 
As already discussed above, this can be seen as a reformulation in the dual domain of the filter-based interpretation of the two properties (smoothness and the exponential decay) included in the Bayesian framework by specifying the TC kernel parametrisation.

By relying on this alternative interpretation, it becomes possible to define the regularisation problem in an intuitive way, by injecting prior information about the system directly at the cost function level. Thanks to this approach, a unified framework is developed to deal quite easily with low-pass, high-pass, band-pass systems, and systems with one or multiple resonances, as will be shown in the next sections.

\section{Solving regularisation problems via the filter-based method}
\label{sec5}


From (\ref{regsolution}) and (\ref{RdecF}) it follows that, once the filter matrix $F$ is defined, one can directly obtain the regularised solution:
\begin{equation}
\label{regsolutionF}
\hat\theta_{\text{reg}}=(\Phi^T\Phi+\lambda F^TF)^{-1}\Phi^TY.
\end{equation}
The formulation of $F$ will typically depend on a number of hyperparameters that need to be optimised by the user, together with the scaling factor $\lambda$, possibly in an automated way.

\subsection{Building the regularisation filter matrix $F$}
\label{sec5a}

To be able to accurately model the impulse response of different type of linear systems, it seems appealing to include in $F$ information about the frequency band of the system. On the other hand, when dealing with stable systems, the decaying nature of the impulse response should also be encoded in $F$. The idea here is to construct $F$ in a flexible way to allow for different system properties, and to tune the hyperparameters in a separate optimisation step to validate these properties on the available input--output data (similar to what is already done in kernel-based regularisation).

Following the intuitive explanation in Section~\ref{sec4}, every row of $F$ should contain filter coefficients that operate on $\theta$. Let us denote by $p$ the order of this regularisation filter, chosen such that $p<n$ holds, to guarantee that the $p+1$ filter coefficients can be included in the $n$-dimensional rows of $F$. This constraint does not constitute a limitation for the algorithm, since $n$ is typically chosen to be large enough.

Given two cut-off frequencies $f_1$ and $f_2$, with $f_1<f_2$, and the filter order $p$, determine the $p+1$ filter coefficients $b_0,\ldots,b_p$, e.g.\ by using the MATLAB \texttt{fir1} function (note that many other ways of designing a suitable filter can be used). Note also that the choice of using two cut-off frequencies allows us to deal in a straightforward way with low-pass, high-pass, and band-pass systems. 

Once the coefficients $b_0,\ldots,b_p$ are obtained, the filter matrix $F$ is built as:
\begin{equation}
F=\begin{pmatrix}
\alpha^{-1/2} b_0 &  \cdots & \alpha^{-1/2} b_p & 0 & \cdots & 0 \\
0 & \alpha^{-1} b_0 & \cdots & \alpha^{-1} b_p & 0 & \vdots \\
\vdots & \ddots & \ddots & \ddots & \ddots & 0 \\
\vdots & \ddots & 0 & \alpha^{-(n-p)/2} b_0 & \cdots & \alpha^{-(n-p)/2} b_p \\
\vdots & \ddots & \ddots & \ddots & \ddots & \vdots \\
0 & \cdots & \cdots & \cdots & 0 & \alpha^{-n/2} b_0
  \end{pmatrix}.
\end{equation}
Note that, to include the exponential decay component, the $i$-th row of $F$ is scaled with $\alpha^{-i/2}$, $0\leq\alpha\leq1$. In this way, filters associated to rows with higher number will have higher gain, i.e.\ $\theta$ values at the tail of the impulse response will be penalised more.


The filter order $p$, the two cut-off frequencies $f_1$ and $f_2$, the decay parameter $\alpha$, and the scaling factor $\lambda$ are hyperparameters that need to be optimised, as explained in the next subsection.

To understand how the filter matrix $F$ can be built in the special case of resonance systems with multiple resonances, the reader is referred to the discussion in Section~\ref{secres}.

\begin{figure}[t]
\centering
\includegraphics[width=7.5cm]{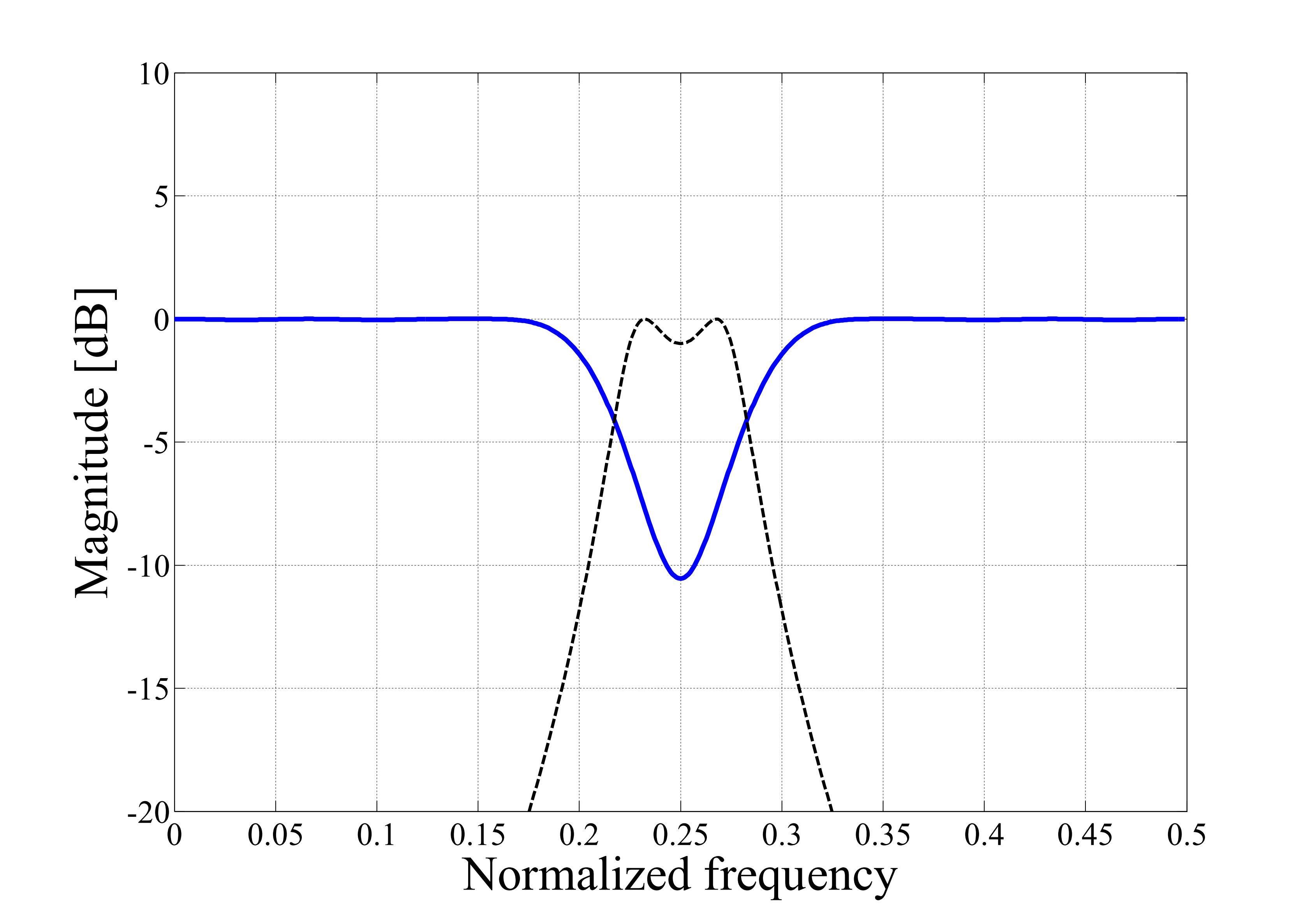}
\caption{Band-pass system modelling example. Magnitude (in dB) of the frequency response of the true system (dashed black line), and of a filter with order 30 used to build matrix $F$ (solid blue line). Note that the filter content in $F$ needs to compensate for the system behavior, since it appears in the cost function as a penalty term.}
\label{figfilt}
\end{figure}

It is important to stress that the characteristic of the filter used to build $F$ needs to be the inverse of the assumed system's behavior. This is due to the fact that the frequency components outside the band of the system need to be penalised in the cost function, while only the frequency components inside the band of the system should be present in the estimated impulse response. An example of this is illustrated in Figure~\ref{figfilt}, where to model a band-pass system a band-stop filter is used to construct $F$.

As a reference, note that the rows of the filter matrices $F_{DC}$ and $F_{TC}$ corresponding to the DC and TC kernels (see eqs.(\ref{DCfiltermat}) and (\ref{TCfiltermat})) represent first order filters ($p=1$) with coefficients determined by the hyperparameters $\rho$ and $\alpha$. The proposed filter-based method allows thus for higher flexibility than the existing approach.

An extension of the stable spline kernels was already introduced in \cite{Pil11}. However, the main difference with respect to that work is that here the definition of the filter matrix is done at the cost function level, while in \cite{Pil11} a finite-dimensional component was added directly to the kernel (i.e.\ the inverse of the regularisation matrix). The reason for this was that in that paper the high-frequency poles were introduced to capture oscillations due to the ARMAX noise model. On the other hand, in the approach presented in this paper the flexibility allowed by the definition of the filter matrix results in a more general method to include in a natural way different system properties in the regularisation approach.

\subsection{Tuning the hyperparameters}
\label{sechyp}

Tuning the hyperparameters that characterise $F$ is a critical step, which can affect the performance of the algorithm. 
The hyperparameters are typically optimised based on the available data, to match the properties of the system as accurately as possible.

In the kernel-based approach, the hyperparameters are tuned on the estimation data by exploiting the Bayesian framework, in particular by using the robust marginal likelihood maximisation method (known as Empirical Bayes) \cite{Chen13}.

However, since in this work an alternative formulation of the estimation problem is considered, the hyperparameters are optimised following a different procedure. 

Let $\beta=[p,f_1,f_2,\alpha,\lambda]$ be the hyperparameter vector (see Section~\ref{sec5a} for the details). 
Here $\beta$ is tuned by minimizing the $k$-fold cross validation mean square error (MSE) \cite{has09}. A simple grid search in the $\beta$ space can be considered for the optimisation, or, alternatively, more sophisticated nonlinear optimisation algorithms can be employed.
One could for instance choose to perform first a fast scan on a coarse grid of values, and then run a nonlinear optimisation algorithm scanning the full hyperparameter space, starting from the initial values found at the previous step. 

The cost function evaluation at each grid search step has computational complexity $O(n^3)$, which is comparable to the complexity required by each step in the nonlinear optimisation routine for the kernel based-methods, see \cite{Chen13}. The total complexity of the filter-based approach in its current implementation depends then on the number of points on the hyperparameter grid for which the cost function is evaluated.

Note that the dimension of the hyperparameter vector for the DC and TC kernels in eqs.~(\ref{DC}) and (\ref{TC}) is 4 and 3, respectively, including $\sigma^2$ that typically also needs to be estimated from data. This means that in the filter-based approach one or two additional hyperparameters need to be estimated. This additional computational load is compensated by the higher flexibility offered by the proposed approach in the design of the filter matrix $F$, while it should be noted that the coefficients in $F_{DC}$ and $F_{TC}$ in eqs.~(\ref{DCfiltermat}) and (\ref{TCfiltermat}) correspond only to a first order filter. Moreover, the direct link of some hyperparameters with physical properties of the system (e.g.\ the cut-off frequencies $f_1$ and $f_2$) makes it easier for the user to tune their values, and to include prior knowledge, if available. 

Note finally that the proposed filter-based method still need to be fine-tuned. Therefore, although the examples tested in the next section suggest that the proposed approach is successful in estimating accurate models, the implementation of the hyperparameter tuning could still be improved. Moreover, the possibility of considering non-causal filters to design the matrix $F$ could also be investigated.

\section{Simulation results}
\label{sec6}



\subsection{Low-pass, band-pass and high-pass systems}

\subsubsection{Settings}
\label{secset}

The following modelling task is considered: based on a set of input--output data, find an impulse response estimate describing the underlying system's behavior as accurately as possible. 

The systems considered in the examples are Chebyshev type 1 filters
$H(z)=\frac{b_0+b_1z^{-1}+\cdots+b_{n_b}z^{-n_b}}{1+a_1z^{-1}+\cdots+a_{n_a}z^{-n_a}}$
with 1 dB peak-to-peak ripple in the passband and frequency band (normalised with respect to the sampling frequency) as detailed below. The coefficient vectors $b$ and $a$ are obtained running the MATLAB command \texttt{cheby1}. The FIR model order $n$ is fixed equal to $100$, which is sufficiently large to approximate the true system response in all cases.

The five different systems under test are:
\begin{itemize}
\item a second order low-pass system with normalised frequency band $[0 \ 0.05]$;
\item three fourth order band-pass systems with different normalised frequency band: $[0.1 \ 0.15]$, $[0.225 \ 0.275]$ and $[0.35 \ 0.4]$;
\item a second order high-pass system with normalised frequency band $[0.45 \ 0.5]$.
\end{itemize}

The excitation signal is a white Gaussian noise sequence (with zero mean and unit standard deviation) of length $N=250$.  The output is corrupted by white Gaussian noise (with zero mean and standard deviation equal to $0.1$), resulting in a typical SNR of 11 dB. A Monte Carlo simulation with 100 different input and noise realisations is performed. At each Monte Carlo run, the procedure explained in Section~\ref{sec5} is used to estimate the filter-based model (\ref{regsolutionF}). The least squares estimate (\ref{LSsol}), and the kernel-based regularised estimates (\ref{regsolution}) with the DC and TC kernels in (\ref{DC}) and (\ref{TC}) are also computed.

The kernel-based solutions are obtained with the \texttt{arxRegul} function with the standard settings in the R2013b version of the MATLAB System Identification Toolbox \cite{SysIdToolbox}.

For the filter-based approach, the hyperparameters are tuned by minimizing the $2$-fold cross validation MSE (i.e.\ using only the available $250$ estimation data), with a grid search procedure to scan different values of the regularisation filter order $p$ (ranging from $2$ to $30$), the normalised cut-off frequencies $f_1$ and $f_2$ (from $0$ to $0.5$), the scaling factor $\lambda$ (from $1$ to $500$), and the decay parameter $\alpha$ (from $0.7$ to $0.9$).

The performance of the estimated models is evaluated in terms of MSE on a very long noiseless validation set ($N_{\text{val}}=10000$) as follows:
\begin{equation}
{\text{MSE}}_{\text{val}}=\frac{1}{N_{\text{val}}}\sum_{t=1}^{N_{\text{val}}} (y_{\text{val}}(t)-\hat{y}_{\text{val}}(t))^2,
\end{equation}
where $\hat{y}_{\text{val}}$ denotes the modelled validation output.

Since the system is excited with white Gaussian noise, this criterion is equivalent to the fit of the estimated impulse response with respect to the true system response, given that
\begin{equation}
E\{{\text{MSE}}_{\text{val}}\}=\sigma_u^2\sum_{k=0}^{n-1} (g_k-\hat{g}_k)^2,
\end{equation}
where $E\{\cdot\}$ is the expected value w.r.t. the validation input and $\sigma_u$ is the validation input standard deviation (in this case equal to 1).

\subsubsection{Results}

\begin{figure}[t]
\centering
\subfigure[Low-pass system.]{
\includegraphics[width=7.5cm]{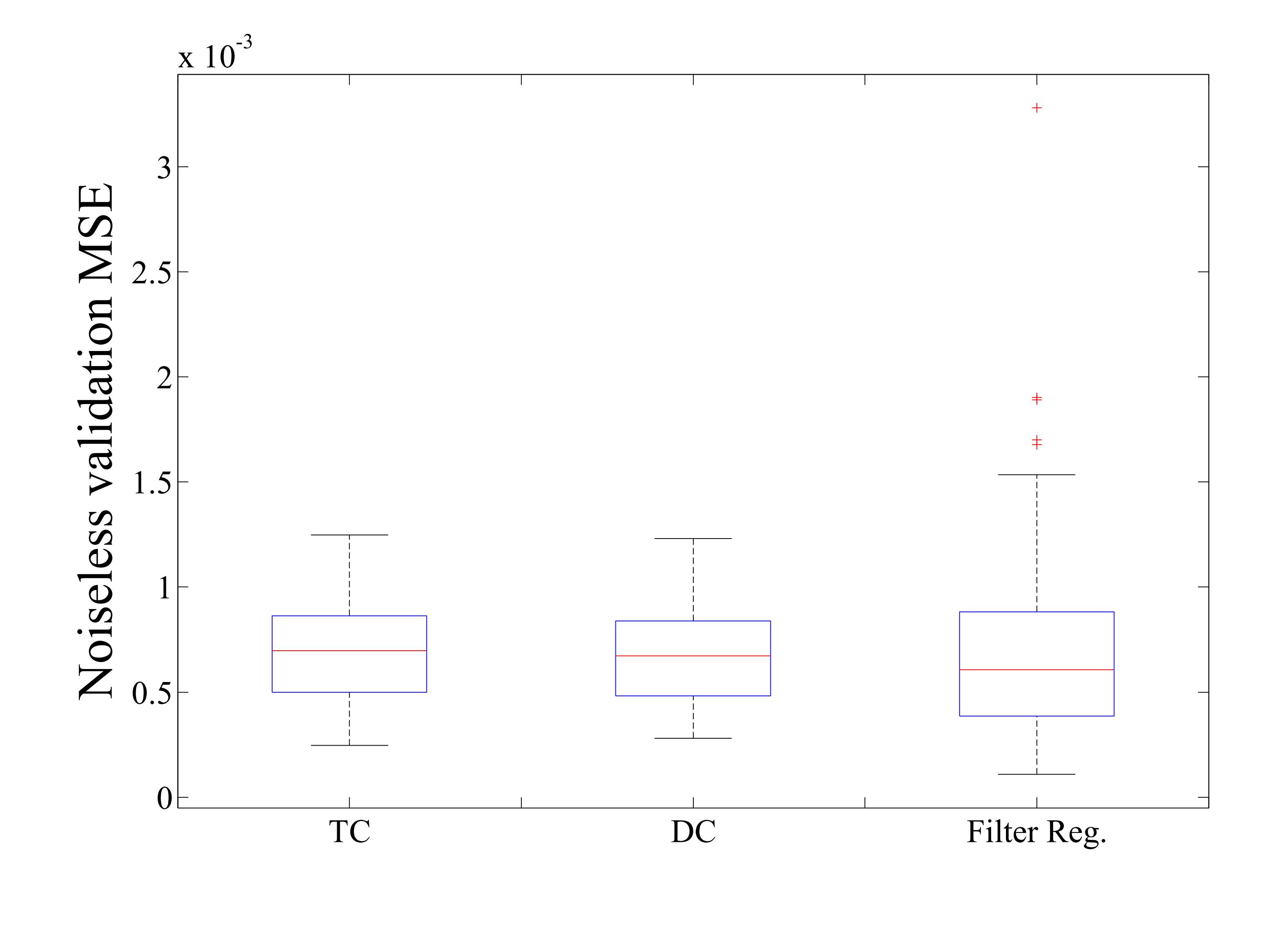}
\label{figbox1}}
\subfigure[High-pass system.]{
\includegraphics[width=7.5cm]{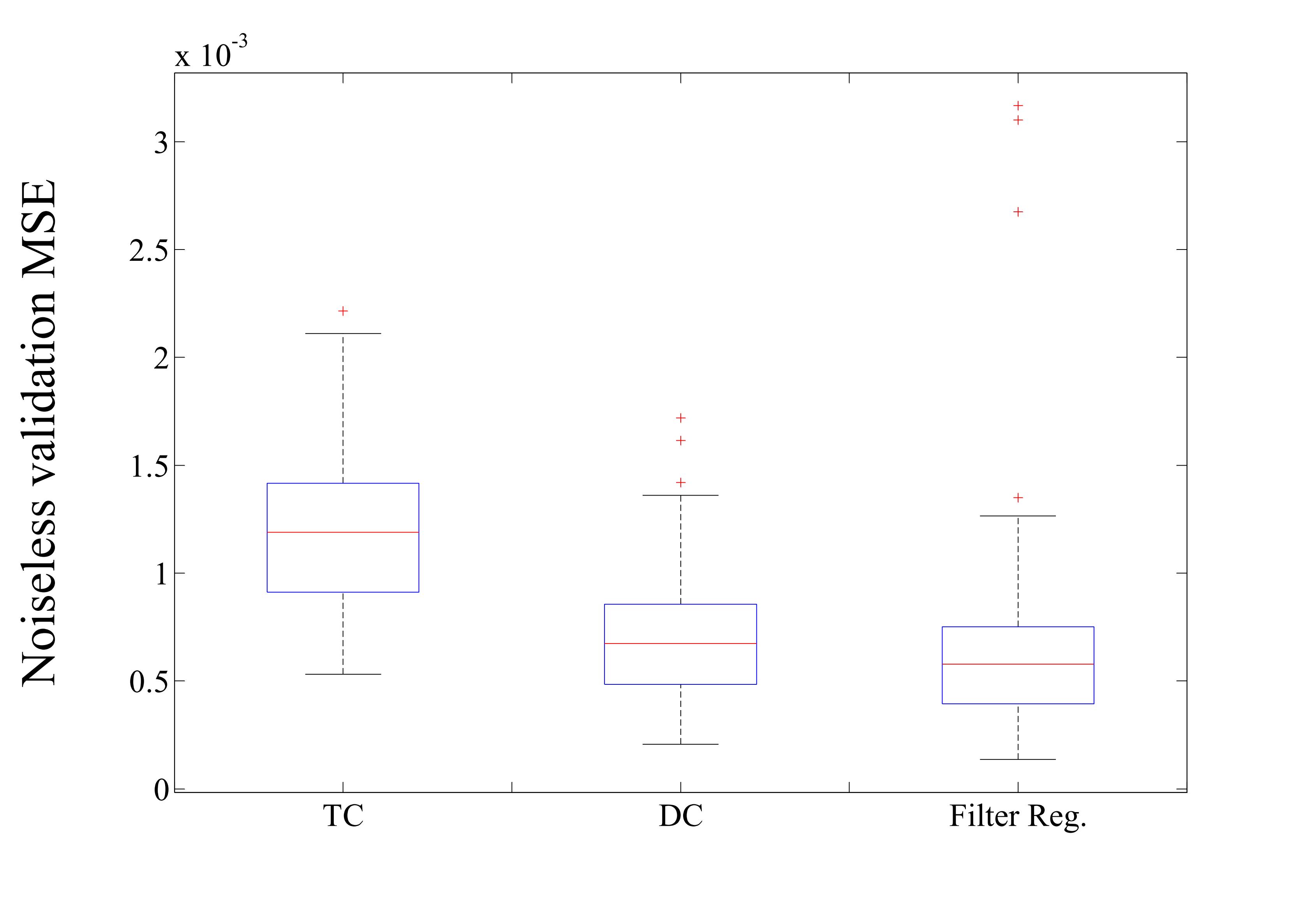}
\label{figbox9}}
\caption{
(a) Low-pass system (normalised frequency band $[0 \ 0.05]$) and (b) high-pass system (normalised frequency band $[0.45 \ 0.5]$) modelling results. 
Comparison of the noiseless validation MSE values ($N_{\text{val}}=10000$) for different methods: kernel-based regularisation with TC kernel, with DC kernel, 
and filter-based approach. For each method, the boxplot of the MSE values for 100 Monte Carlo realisations is shown.
}
\label{figboxlh}
\end{figure}
\begin{figure}[t]
\centering
\includegraphics[width=8.3cm]{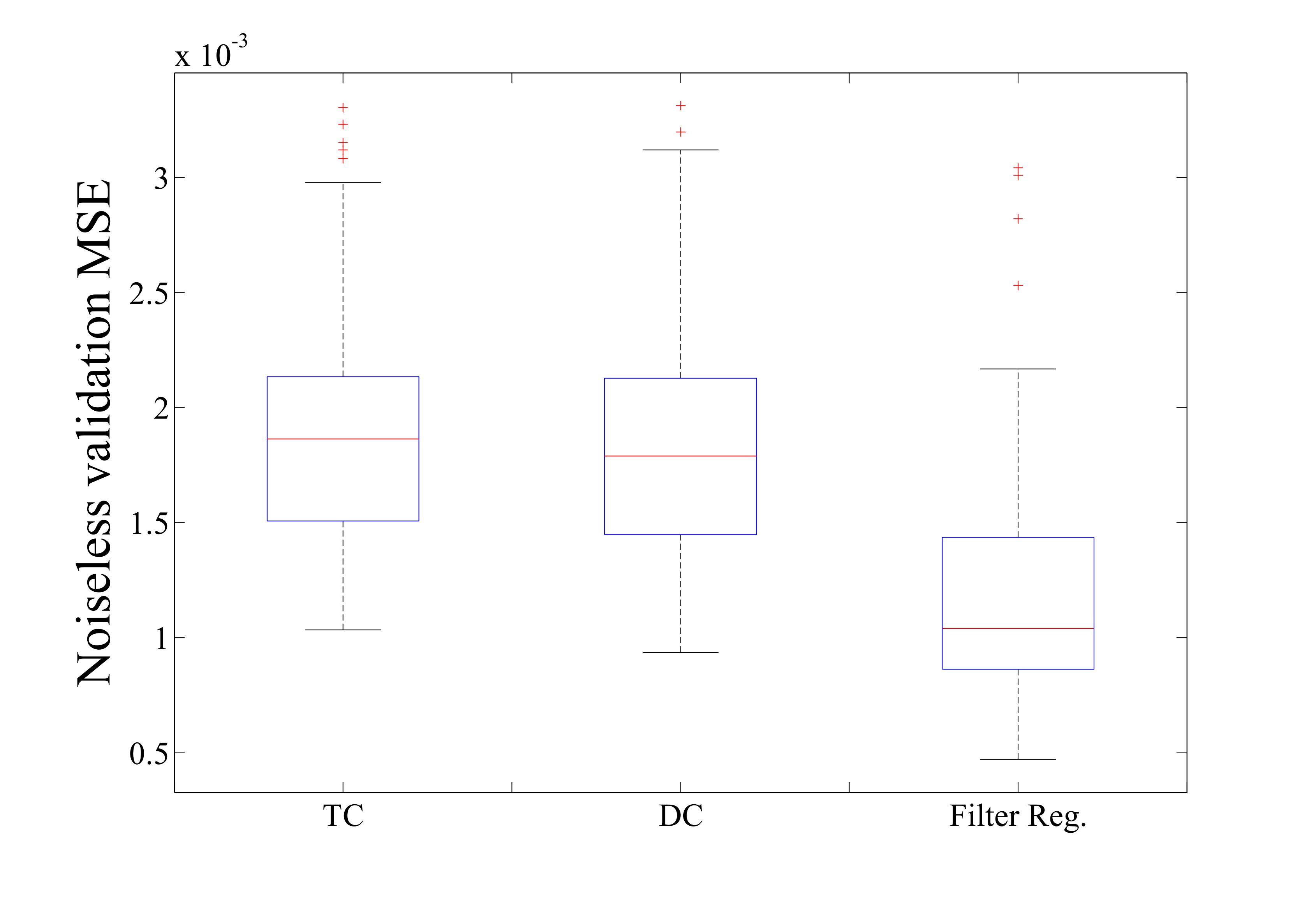}
\includegraphics[width=8.3cm]{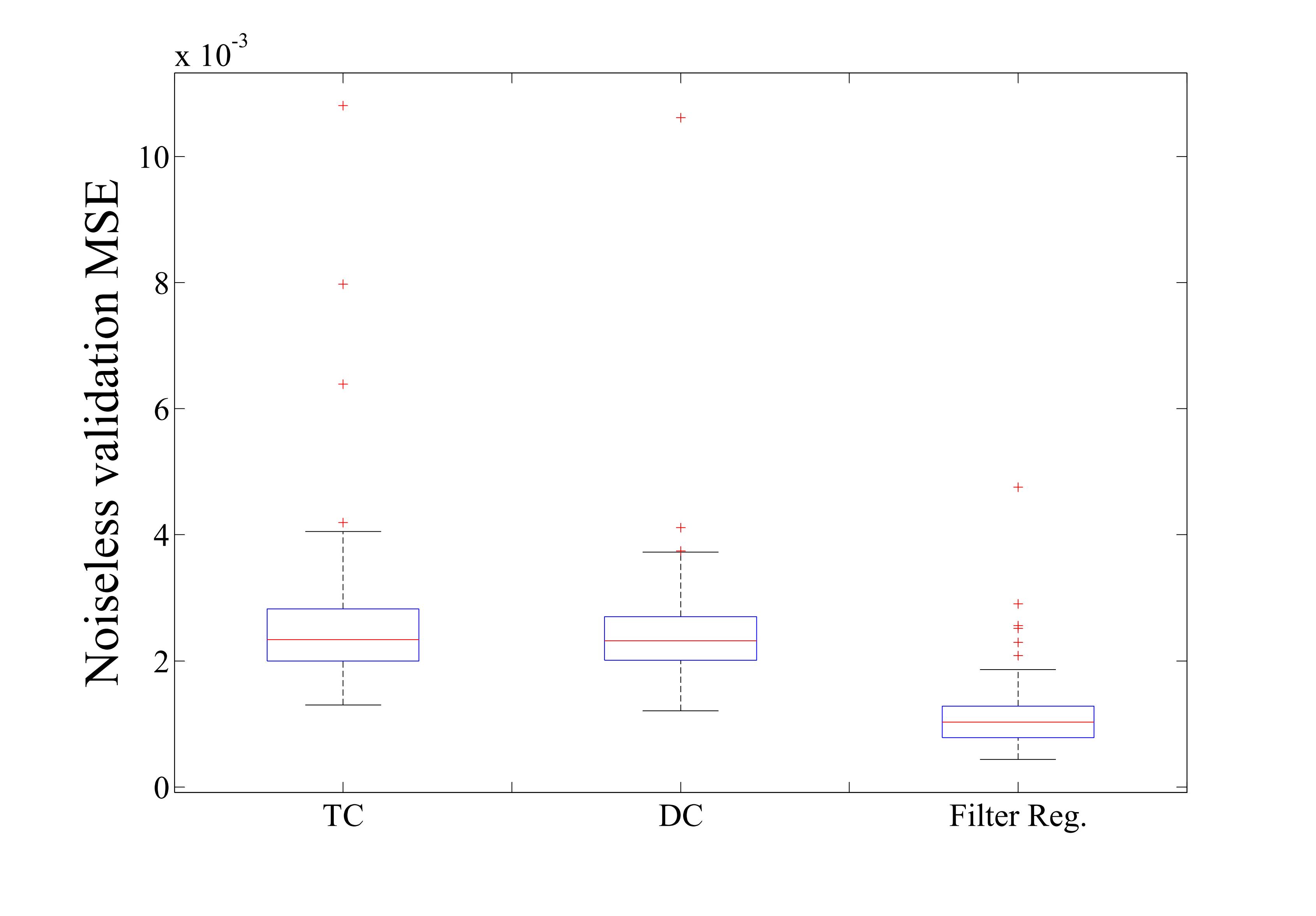}
\includegraphics[width=8.3cm]{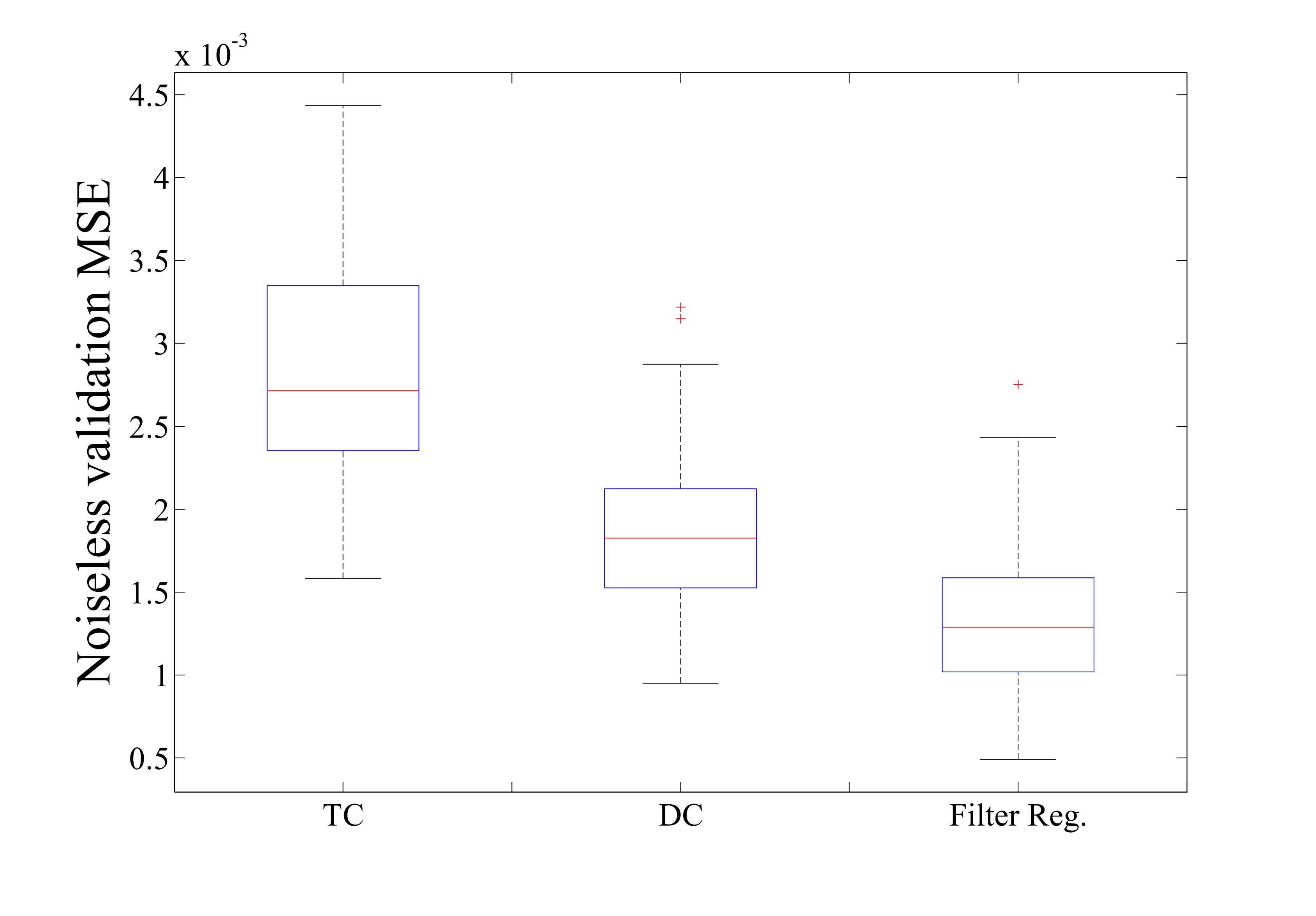}
\caption{
Band-pass system modelling results (normalised frequency band $[0.1 \ 0.15]$ (top), $[0.225 \ 0.275]$ (middle), $[0.35 \ 0.4]$ (bottom)). 
Comparison of the noiseless validation MSE values ($N_{\text{val}}=10000$) for different methods: kernel-based regularisation with TC kernel, with DC kernel, 
and filter-based approach. For each method, the boxplot of the MSE values for 100 Monte Carlo realisations is shown.
}
\label{figboxband}
\end{figure}

Figures~\ref{figboxlh} and \ref{figboxband} show a comparison between the performance of the models estimated by means of the proposed filter-based regularisation approach, and the results of the kernel-based methods. 

All results obtained with the standard least squares approach are much worse than for the regularised solutions, and are therefore omitted to improve the readability of the figures. More precisely, the median of the MSE values for the least squares solution is in all cases between 0.008 and 0.009, which is 7 to 15 times higher than for the proposed approach. 

%
%

The filter-based approach yields better results (up to two times lower values for the median MSE) than the TC and the DC kernel regularisation in all the considered examples. This can be observed from the boxplots of the MSE values, and can also be appreciated by considering the results for the single Monte Carlo runs. Table~\ref{tabperc} reports the percentage values of times in which the filter-based approach outperforms the TC and DC kernels. Note that the results obtained with the TC and DC kernels can still be considered satisfactory since they yield a considerable improvement when compared to the least squares solution. However, the gain in performance given by the filter-based regularisation reflects the flexibility of this approach in dealing with a variety of different systems.

\begin{table}[!t]
\begin{center}
\caption{\label{tabperc} {Percentage values of the amount of Monte Carlo runs in which the proposed filter-based approach outperforms (lower noiseless validation MSE) the TC and the DC kernel-based solutions.}}
\begin{tabular}{ c | c  c  c  c  c }
Kernel & Low & Band 1  & Band 2 & Band 3 & High \\
\hline
TC & 70 & 95 & 97 & 98 & 94 \\
DC & 67 & 94 & 98 & 91 & 76 \\
\hline
\end{tabular}
\end{center}
\end{table}

The filter-based regularisation gives particularly accurate estimates of the impulse response in the band-pass and high-pass examples, but the results are satisfactory also in the low-pass case.

Moreover, it is observed that the selected cut-off frequencies $f_1$ and $f_2$ correspond to the system frequency band in the different examples. The selected regularisation filter order $p$ also guarantees the appropriate gain in the band of interest.

These promising results might even be further improved by implementing a more sophisticated hyperparameter tuning and filter design strategy, as mentioned in Section~\ref{sechyp}. Note that the (very few) bad outliers in the boxplot of the errors obtained with the filter-based approach in the low-pass and in the high-pass cases are due to a wrong choice of the hyperparameters, and could therefore be avoided with an improved hyperparameter optimisation.

\subsection{Resonance systems}
\label{secres}

\subsubsection{Settings}

For the resonance systems examples, the settings are as detailed in Section~\ref{secset}. A few differences are listed below.

Three different systems are considered:
\begin{itemize}
\item one resonance: a second order resonance system with normalised frequency band $[0.145 \ 0.15]$;
\item two resonances (one dominant): sum of two second order resonance systems with normalised frequency bands $[0.145 \ 0.15]$ and $[0.395 \ 0.4]$ respectively. The amplitude of the first resonance system is scaled with a factor $0.2$;
\item two resonances: sum of two second order resonance systems with normalised frequency bands $[0.145 \ 0.15]$ and $[0.395 \ 0.4]$ respectively (equal amplitude).
\end{itemize}

For an illustration of the frequency characteristics, the magnitude of the three considered resonance systems is plotted in the left plots of Figure~\ref{figres}.

During the grid search procedure for the tuning of the hyperparameters, the same values for $p$, $f_1$, $f_2$ and $\lambda$ as in Section~\ref{secset} are scanned, while values of the decay parameter $\alpha$ are considered in the range $0.85$ to $0.95$.

\subsubsection{Results}

\begin{figure}[t]
\centering
\includegraphics[width=6.2cm]{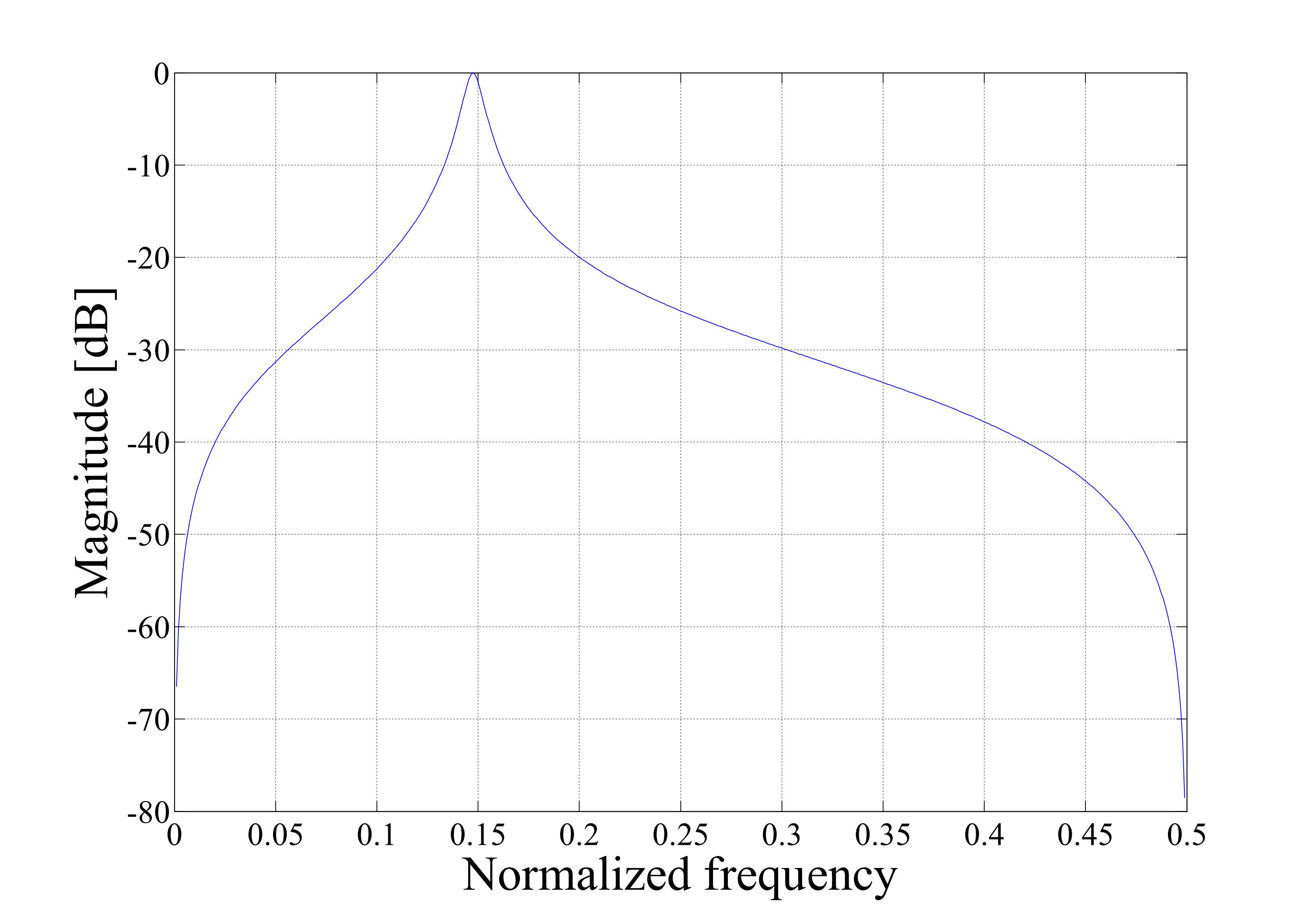}
\includegraphics[width=6.2cm]{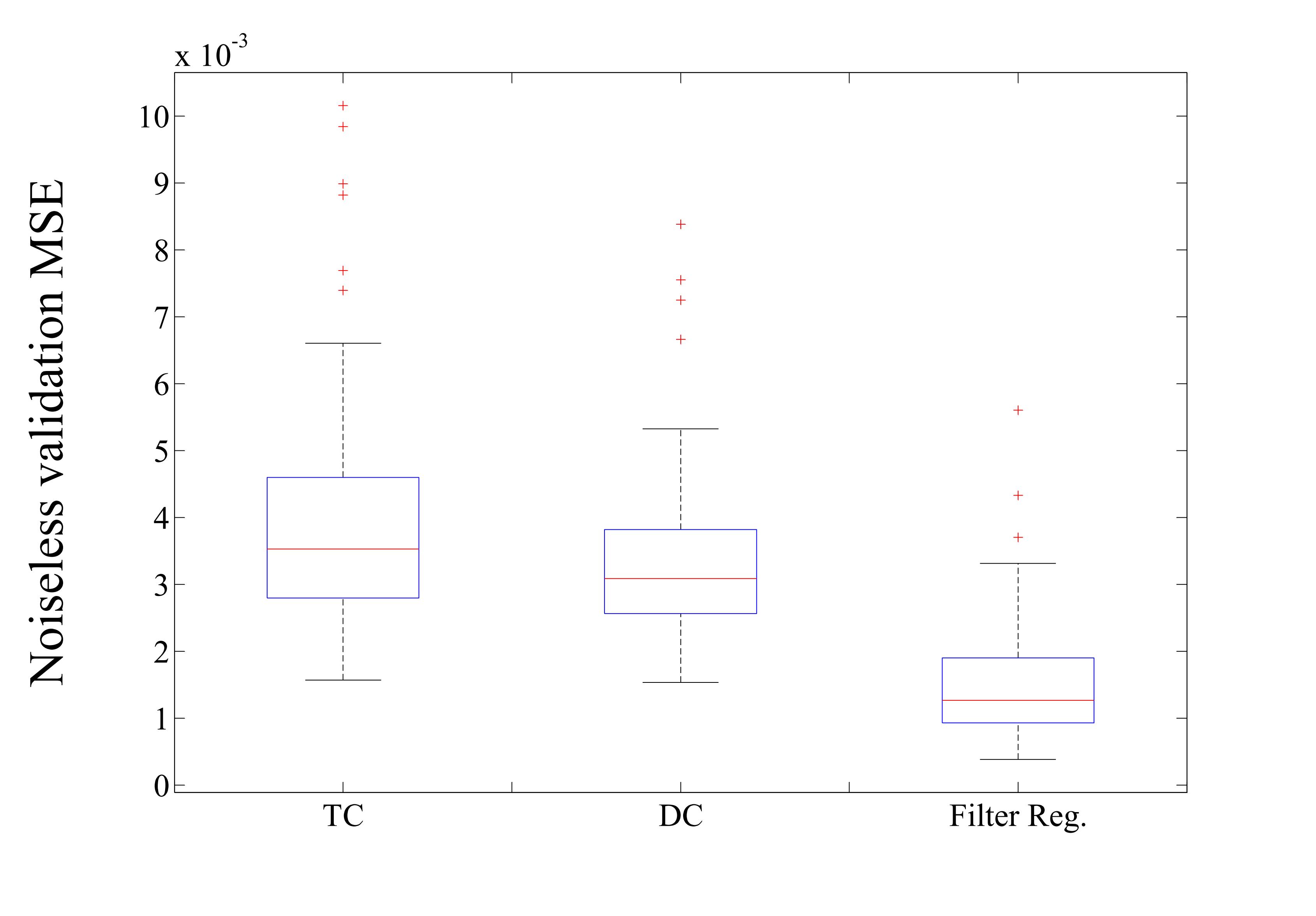}
\includegraphics[width=6.2cm]{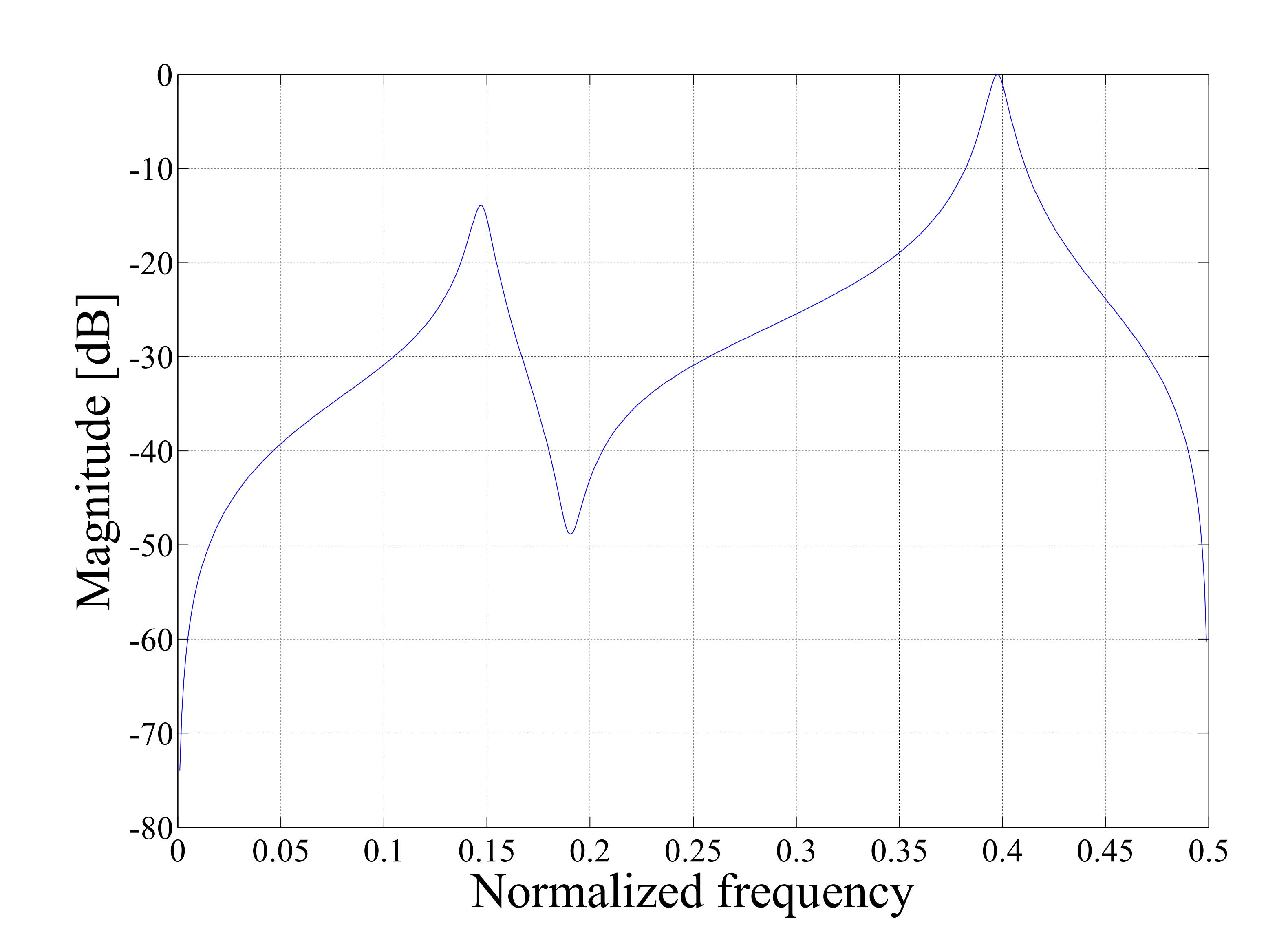}
\includegraphics[width=6.2cm]{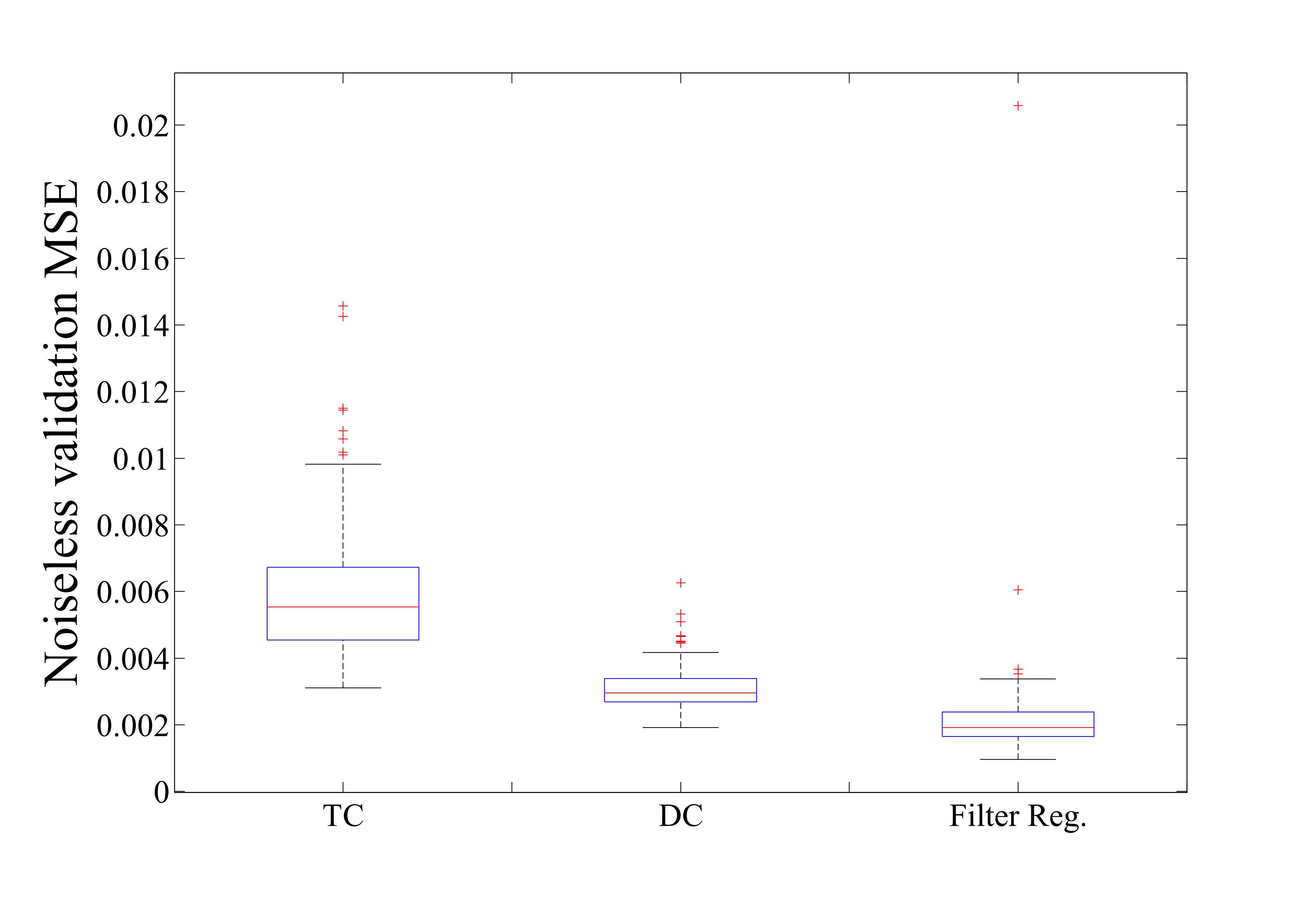}
\includegraphics[width=6.2cm]{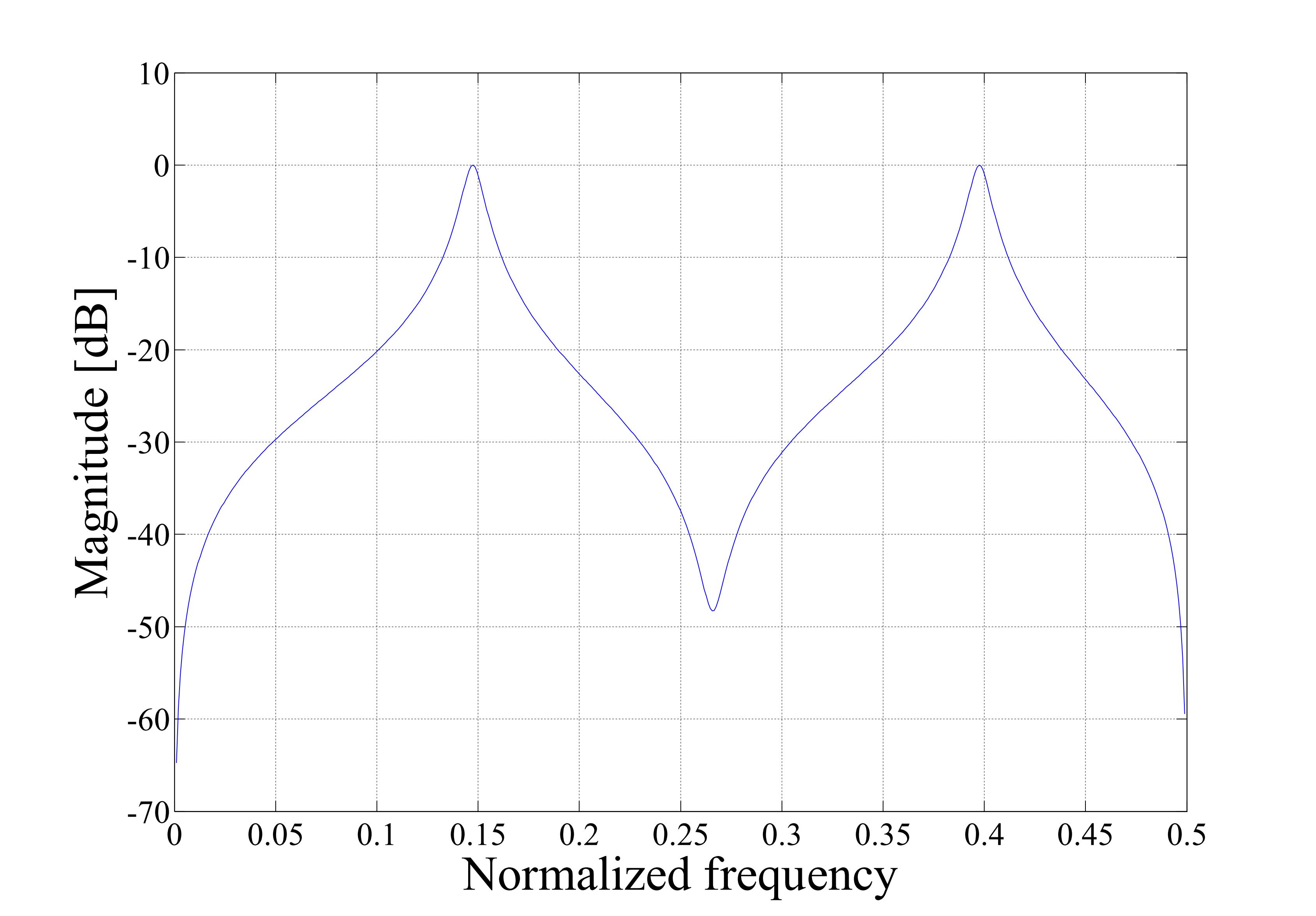}
\includegraphics[width=6.2cm]{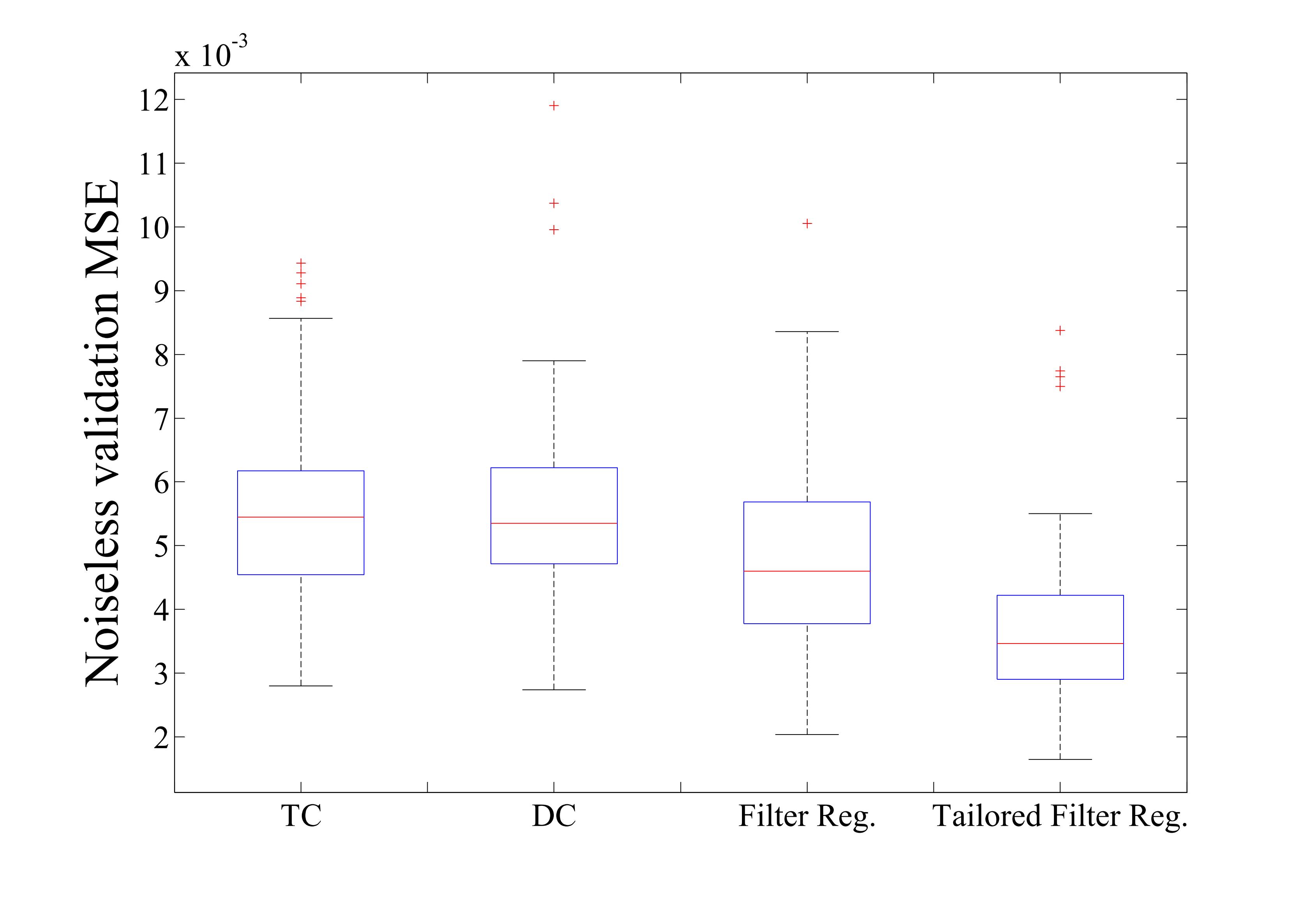}
\caption{Resonance system examples. Top: system with one resonance. Middle: system with two resonances, one dominant. Bottom: system with two resonances. Left: Magnitude (in dB) of the considered system. Right: Comparison of the noiseless validation MSE values ($N_{\text{val}}=10000$) for different methods: kernel-based regularisation with TC kernel, with DC kernel, and filter-based approach (bottom right plot: filter-based approach, without and with a tailored filter design). For each method, the boxplot of the MSE values for 100 Monte Carlo realisations is shown.}
\label{figres}
\end{figure}

The results obtained with the filter-based regularisation on the resonance systems identification examples are shown in Figure~\ref{figres}, and compared with the performance of the TC and DC kernel methods. 

In the first two cases (one resonance, and two resonances of which one dominant) the filter-based approach clearly outperforms the kernel-based regularisation. 

The third example (two resonances, equal amplitude) is much more challenging, since the original version of the algorithm does not allow one to select multiple frequency bands (only two cut-off frequencies $f_1$ and $f_2$ are considered in the hyperparameter set). However, even in this case, a small improvement of the standard filter-based approach can be appreciated in the bottom right plot in Figure~\ref{figres}, in comparison with the TC and DC kernel methods. 

In order to obtain a significant decrease of the error values also in this third example, the algorithm has been modified as follows: instead of scanning different values of $f_1$ and $f_2$ and select the optimal ones to build the filter matrix $F$, a `tailored' filter (with as many frequency bands as needed) is designed, and its coefficients are used in the rows of $F$. This can be realised e.g.\ using information from the output spectrum, which gives some ideas about the characteristics of the system, at least in the case of a white input signal. The filter order $p$ can still be tuned during the grid search selection, together with the other hyperparameters. 

This results in the fourth boxplot in the bottom right plot of Figure~\ref{figres} (tailored filter regularisation). The modified version of the filter-based regularisation is of course less flexible than the original approach, and requires additional prior knowledge from the user. However, in the examples we considered, it turned out that an approximate guess of the system characteristics is sufficient to obtain very accurate models. More in details, for the two resonance system in the bottom left plot of Figure~\ref{figres}, it was sufficient to build the filter by imposing a band-stop behaviour in the normalised frequency bands $[0.1 \ 0.2]$ and $[0.35 \ 0.45]$, i.e.\ the inverse of the assumed system's behavior.

The results on this last example illustrate once more one of the main ideas behind the proposed approach, namely the possibility to include in the identification problem any prior knowledge about the system properties in an intuitive way from an engineering point of view.

\section{Conclusions}
\label{concl}

In this paper, regularisation methods for impulse response modelling are studied from an alternative perspective. The filter-interpretation ideas presented in this work allow one to get more insight about the existing kernel techniques from an engineering point of view. Moreover, they are exploited to design a new user-friendly filter-based regularisation method, by including prior knowledge about the system's properties directly at the cost function level.

The effectiveness of the proposed approach is illustrated by means of Monte Carlo simulations on different modelling examples. The filter-based approach outperforms the standard least squares method and the existing kernel-based regularisation approaches in all the considered examples, and establishes a unified framework to deal in an intuitive way with low-pass, band-pass, high-pass systems, and resonance systems.

Future research steps include the improvement of the hyperparameter tuning procedure, which represents a crucial step in the estimation. Efforts in this direction could lead to even better performance of the filter-based regularisation approach. Moreover, the filter design step could also be improved, e.g.\ by considering, in addition to what done so far, the possibility to use non-causal filters to build the matrix $F$.

\section*{Acknowledgment}

This work was supported in part by the Fund for Scientific Research (FWO-Vlaanderen), by the Flemish Government (Methusalem), by the Belgian Government through the Inter university Poles of Attraction (IAP VII) Program, and by the the ERC Advanced Grant SNL-SID, under contract 320378.

\bibliographystyle{IEEEtran}
\bibliography{bibtex}

\end{document}